\documentclass[epj]{svjour}
\usepackage{graphics}
\usepackage{amsmath,amssymb,bm}
\usepackage{epsfig}
\usepackage{amsfonts}
\usepackage{epstopdf}
\usepackage[usenames,dvipsnames]{color} 
\usepackage{float}

\newcommand{\beqa}{\begin{eqnarray}}
\newcommand{\eeqa}{\end{eqnarray}}
\newcommand{\bd}[1]{ \mbox{\boldmath $#1$}}
\begin{document}
\def\ii{\'\i}
\title{A Semimicroscopic Algebraic Cluster Model
for Heavy Nuclei
}
\subtitle{I: One heavy and one light cluster
}
\author{Peter O. Hess\inst{1,2} \and 
Leonardo J. Ch\'avez-Nu\~nez\inst{1}
}                     
%
%
\institute{Instituto de Ciencias Nucleares, Universidad Nacional 
Aut\'onoma de M\'exico, Circuito Exterior, C.U., 
A.P. 70-543, 04510 M\'exico D.F., Mexico \and Frankfurt Institute for Advanced Studies, Johann Wolfgang 
Goethe Universit\"at, Ruth-Moufang-Str. 1, 60438 Frankfurt am Main, Germany}
\date{Received: date / Revised version: date}
%
\abstract{
An extension of the {\it Semimicroscopic Algebraic
Cluster Model} (SACM) is proposed, based on the
{\it pseudo-}$SU(3)$ model ($\widetilde{SU}(3)$).
The Hamiltonian and the spectroscopic factor operator 
of the model are presented and a procedure of constructing
the model space. Because a huge number of $SU(3)$ irreducible
representations (irrep) appear, one has to be careful
in designing a practical, 
consistent path to reduce the Hilbert
space. The concept of {\it forbiddenness},
taking into account excitations of the clusters,
is introduced and applied. The applications are 
to two systems with a low forbiddenness, namely to
$^{236}$U $\rightarrow$ $^{210}$Pb + $^{26}$Ne and
$^{224}$Ra $\rightarrow$ $^{210}$Pb + $^{14}$C, and
to $^{236}$U $\rightarrow$ $^{146}$Xe + $^{90}$Sr, which appears in the fission of $^{236}U$, which
requires a large forbiddenness.
Energies, electromagnetic transitions and spectroscopic 
factors are calculated.
} 
\maketitle

\section{Introduction}
\label{intro}

The {\it Semimicroscopic Algebraic Cluster Model} (SACM) was proposed in
\cite{cseh-letter,cseh-levai-anph} for 
light nuclei (see, for example, \cite{sacm-appl1}),
with the intend as an alternative to also describe nuclear molecules, i.e., an alegraic version of their geometrical
description \cite{scheid1995,hess-1984}.  
A first attempt to extend it to heavy nuclei is published in 
\cite{cseh-algora}, using the 
$\widetilde{SU}(3)$ (pseudo-$SU(3)$) model \cite{hecht,arima}.
While the heavy cluster is treated within the 
$\widetilde{SU}(3)$ the light cluster is described
by the $SU(3)$ standard model.
In \cite{cseh-algora,cseh-scheid}, 
the separation of nucleons into the unique and normal orbitals in the united 
nucleus, compared to the ones in each cluster, is not well defined:
Distinction is made when a cluster is light, which is then treated within the
standard shell model, or both are heavy. In different words: Each cluster and the united nucleus
are not treated in the same mean field. 

Since then,
many applications of the SACM have been studied and 
further attempts
to extend it to heavy nuclei: For example, a construction of the effective $SU(3)$ irreducible
representations (irreps)  for heavy nuclei \cite{hunyadi}, using the 
Nilsson model \cite{ring}, and a study of preferences in radioactive decays and/or fission
\cite{sacm-fission1,sacm-fission2,sacm-fission3}.   
In \cite{hess-86}, the 
spectra of
$\alpha$ cluster nuclei, of significant interest in astrophysics related to the production  
of heavy elements, and their spectroscopic factors were calculated. In \cite{phase-I,phase-II}
the first steps in investigating phase transitions
were taken and more recently 
\cite{david} a complete description of phase transitions 
using the catastrophe theory \cite{gilmore} was published.
In \cite{renorm}
the renormalization of the coherent state parameters
is investigated, used for the geometric
mapping. 

More recently, the $\widetilde{SU}(3)$ model
was applied to the {\it shell like quarteting in
heavy nuclei} \cite{cseh2020}, another method
to restrict effectively the shell model space
on physical grounds. 
The quarteting model was first proposed in
\cite{arima-quart,danos-quart} and applied
in \cite{cseh-quart} within the SACM for light nuclei. However, the proton and the neutron part are
coupled directly to a total $SU(3)$ {\it irreducible
representation} (irrep) without taking into account the
preference for certain couplings, leading 
as a result to too many irreps
at low energy. This is also a problem of the model we 
present in this contribution and a
path on how to tackle it is proposed.
The quarteting model was
compared to another procedure, called the {\it proxy}-$SU(3)$
\cite{bonatsos2017}, showing comparable results.
All methods have in common to exploit a symmetry,
showing up in the single particle spectrum, and  using
it to effectively cut the model space
to a manageable size. Here we will
propose an equivalent manner to describe heavy nuclei,
using the $\widetilde{SU}(3)$ model and a proposal on how to restrict the model space to the most important contributions.

A further prove of the success of the SACM is the
introduction of the {\it multi-channel symmetry} \cite{multi},
were different clusterizations were connected via the same
Hamiltonian, thus, reducing the complexity of the model and
delivering more insight into the structure of cluster systems.

In \cite{hunyadi} a quite powerful method was presented on 
how to treat heavy
nuclei, it only delivers the {\it ground state}, or the first super- and hyper-deformed states
\cite{cseh2006}. Therefore, it is of interest to
look for alternative procedures  in order to deal also with excited states. Unfortunately, we cannot list all contributions of the SACM, but the ones cited clearly demonstrate the success of the SACM. 

The $\widetilde{SU}(3)$ was proposed in \cite{hecht,arima}
for its application to heavy nuclei, where it is
observed that when
the intruder states, called {\it unique orbitals}, 
with $j = \eta+\frac{1}{2}$ in each
shell ($\eta$ is the shell number), are excluded, the
remaining orbitals, called {\it normal orbitals}, are grouped, just by counting, into
shells of what is denominated as the $\widetilde{SU}(3)$
symmetry. The normal orbital in the $\eta$-shell are renamed
(see next section) such that they correspond to a
pseudo-shell of ${\tilde \eta}=\eta -1$.
Within the Nilsson model asymptotic states also show 
a degeneration in orbitals denoted by the asymptotic
quantum numbers of the Nilsson model
$\Omega\left[\eta \eta_z\Lambda\right]$ 
\cite{ring,plb1994}.

That the nuclear force exhibits
such an unexpected symmetry is today well understood,
parting from microscopic field theoretic models of 
the nuclear interaction \cite{fieldsu3} 
and mapping to the effective nuclear
interaction. The complete Hilbert space of the shell model 
is a direct 
product of a state described within a $\widetilde{SU}(3)$,
in the same manner as the $SU(3)$ model of Elliott
for light nuclei, and a state describing the nucleons in
the intruder orbitals. The nucleons in the intruder orbitals 
are assumed to play only the role of an observer: Nucleons
in the unique orbitals play a passive role
in the dynamics due to their opposite parity and their
coupling to spin-zero pairs, 
contributing only to the binding energy. The pairing energy
of nucleons in the unique orbitals is big, due
to the large $j$ \cite{ring}, compared to 
the nucleons in the normal
orbitals. Thus, nucleons in the unique orbitals only
contribute at high energies, e.g., in the back-bending 
effect and related phenomena \cite{draayer-book}.  
Inter-shell excitations are considered only within the
normal states, for the same reasons.
The contribution of the nucleons in the unique orbitals
are treated via well defined effective charges for electromagnetic transitions \cite{NPA1994}. The effective
charges do not depend on parameters, but only on the total number
of nucleons $A$, ${\tilde A}$ and protons $Z$, ${\tilde Z}$,
where the symbols with the tilde refer to numbers in
the normal orbitals.
In conclusion, the restriction to the $\widetilde{SU}(3)$
is well justified for states at low energy.
Though, the $\widetilde{SU}(3)$ has its limits, as not including independent
dynamical effects of nucleons in the unique orbitals, it is still useful for investigating
an extension of the SACM to heavy nuclei.

Already within the SACM for light nuclei,
new problems arise: When the lightest
cluster is too large and assuming both clusters in their ground state, the corresponding
$SU(3)$ irrep 
of the united nucleus cannot be reached. The main reason is that for increasing number of
nucleons of the light cluster, the number of quanta
in the relative motion becomes too large and the coupling of the relative motion irrep to the
cluster irreps leads to large final $SU(3)$ irreps,
not coinciding wityh the ground state irrep of the
united nucleus.
This problem was recognized in
\cite{smirnov} and the authors introduced the concept of {\it forbiddenness}. The main idea is as follows:
When the two clusters are in their ground state and all 
missing quanta are put into the relative motion 
(the Wildermuth condition \cite{wildermuth}) and
the ground state of the united nucleus 
can be reached, everything is fine. No excitations
are then allowed within a cluster, because these are included
in the excitation of the relative motion. However, when the 
ground state cannot be reached, one 
or the two clusters are allowed to be
excited, subtracting quanta from the relative motion, 
up to the point when for the first time
the ground state is reached.
The excited state of the cluster ($C_k^*$) is then fixed 
and the rest of the procedure is identical as in the SACM
for light nuclei.  
The minimal number required for the excitation of the clusters 
is called the {\it forbiddenness}. 

In \cite{smirnov} the importance of this concept 
and its consequences was proven and applied quite successfully in the 
prediction of preferences 
of fission fragments and fusion of light and heavy systems. The larger the {\it forbiddenness}, 
the more suppressed is the reaction channel.
To summarize: For low lying states and a
large light cluster, the cluster system has to be excited!
Reconsidering the definition of {\it forbiddenness} from an alternative angle, we were able to 
obtain a simple formula on the minimal number of 
excitation quanta \cite{huitz-2015} needed, which will
be resumed further below.

Both models, the $\widetilde{SU}(3)$ and the 
SACM extended to heavy nuclei, will be explained 
in section \ref{sacm-ex}. The formalism includes
a restricted model Hamiltonian,
the calculation of spectra, electromagnetic transition
rates and the determination of spectroscopic factors.
In particular, in Section \ref{examples} we will demonstrate  
the importance of the concept of the {\it forbiddenness}.
In Section \ref{examples} the model is applied to
several heavy cluster systems. 
Two systems only require a low number
of forbiddenness, while the third system
exhibits a large forbiddenness and demonstrates 
its importance.
In section \ref{conclusions} conclusions are drawn.

\section{The SACM and its extension to heavy nuclei}
\label{sacm-ex}

The approach to $\widetilde{SU}(3)$ is as follows:
In each harmonic oscillator shell $\eta$
the orbital belonging to the
largest spin $j=\eta + \frac{1}{2}$ is removed from
consideration as an active orbital, it is considered
rather as a spectator.
To the remaining orbitals the redefinition

\beqa
j ~=~ l\pm\frac{1}{2} & \rightarrow & {\tilde l}~=~ 
l\mp\frac{1}{2}
~,~
\eta ~ \rightarrow ~ {\tilde \eta} ~=~ \eta - 1
~~~,
\label{eq-1}
\eeqa
is applied, where ${\tilde l}$ denotes the
{\it pseudo-orbital angular momentum}
and ${\tilde \eta}$ the pseudo-shell number. With
this redefinition alone it is easily verified that
each shell $\widetilde{\eta}$ has the same content
as the corresponding shell in the standard $SU(3)$ 
model.

For large deformations, the Nilsson states for axial symmetric 
nuclei are classified by their asymptotic quantum numbers
\cite{ring}

\beqa
\Omega \left[ \eta\eta_z\Lambda \right]
~~~,
\label{nil-1}
\eeqa
where $\eta_z$ is the oscillation number 
in the $z$-direction, $\Lambda$ is the projection 
of the orbital
angular momentum onto the same axis and $\Omega$
= $\Lambda \pm \frac{1}{2}$.  

Excluding the intruder levels, the 
reassignment of the orbitals is

\beqa
\Omega ~=~ \Lambda \pm\frac{1}{2} & \rightarrow & 
\widetilde{\Lambda}~=~ 
\Lambda\mp\frac{1}{2}
~~~.
\label{eq-1a}
\eeqa
Inspecting the Nilsson diagrams \cite{ring},
those orbitals with {\it the same quantum numbers}
$\left[\widetilde{\eta}\widetilde{\eta_z}
\widetilde{\Lambda}\right]$ are degenerate,
which implies a very small pseudo-spin-orbit interaction
and as a consequence an {\it approximate symmetry}. In addition, the content
of the $\widetilde{\eta}$ shell corresponds to the 
same one in the
standard shell model.
Thus, the shell model for light
nuclei can be directly extended to heavy nuclei, 
using the $\widetilde{SU}(3)$ model instead of the $SU(3)$ model. 

The basis in the Hilbert space is a direct product
of the $\widetilde{SU}(3)$ states and the ones describing
the unique (intruder) orbitals. 
As mentioned in the introduction, their contribution to the
nuclear dynamics is taken into account by 
a well defined scaling factor (effective charge).

For example, the quadrupole effective charge is given by
\cite{NPA1994}:

\beqa
e_{eff}(E2) & = & \left(\frac{Z}{\widetilde{A}}\right)^2
\left( \frac{A}{{\widetilde A}}\right)^{\frac{4}{3}}
~~~,
\label{nil-2}
\eeqa
where $A$ and ${\widetilde A}$ 
are the total number of nucleons
and the number of nucleons in the normal orbitals,
respectively. $Z$ is the total number of protons.

Restricting the dynamics only to the nucleons in the
normal orbitals was justified in the introduction,
recognizing that at large excitation energies,
where backbending effects play a role, the model
has to be modified by adding the
contributions of the dynamics in the unique orbitals.
The effectiveness of $\widetilde{SU}(3)$ was
demonstrated in \cite{NPA1994}, for the case of
the pseudo-symplectic model of the nucleus
\cite{pseudo-sympl} and many other applications of the 
$\widetilde{SU}(3)$ model (see, for example, 
\cite{cseh-quart}).

\subsection{Construction of the model space in
light nuclei}

In the SACM for 
{\it light nuclei}, 
the $SU(3)$ irreps are determined using the following path:
Each cluster is represented by an irrep $\left( \lambda_k,\mu_k\right)$ ($k=1,2$)
{\it in their ground state}. 
Adding the number of oscillation quanta 
contained in each cluster and
comparing them with the number of oscillation quanta 
of the united nucleus results in
a mismatch: The number of oscillation quanta of the united nucleus is larger than
the sum of both clusters. Wildermuth \cite{wildermuth} showed that the necessary condition
to satisfy the Pauli exclusion principle is to add 
the missing quanta into the relative motion, 
introducing a minimal number of relative 
oscillation quanta $n_0$.
This is known as the {\it Wildermuth condition}.  However, there are 
still irreps which are not allowed by the Pauli-exclusion principle. 

An elegant solution to it, avoiding cumbersome {\it explicit} antisymmetrization of the wave function, was proposed in \cite{cseh-letter,cseh-levai-anph}, 
the original publication of the SACM: 
The coupling of the cluster irreps with the one of the relative motion generates a list of $SU(3)$ irreps, i.e.,

\beqa
\left(\lambda_1,\mu_1\right) \otimes \left(\lambda_2,\mu_2\right) \otimes
\left(n_\pi , 0\right) & = & \sum_{m_{\lambda , \mu} } m_{\lambda , \mu} 
\left( \lambda , \mu \right)
~~~,
\label{eq-2}
\eeqa
where $n_\pi$ is the number of relative oscillation quanta
($\pi$ gives reference to the relative $\pi$-bosons of 
spin 1), limited from below
by $n_0$, and $m_{\lambda , \mu}$ is the multiplicity of
$\left( \lambda , \mu \right)$.
This list of irreps is compared to the one of the shell model. Only those, which have a counterpart in the shell model,
are included in the SACM model space.
In this manner, the Pauli exclusion principle is observed and the model space can be  called  microscopic.

In such a manner, the basis states are described by the
ket state

\beqa
\mid (\lambda_1,\mu_1)(\lambda_2,\mu_2); 
\rho_C (\lambda_C,\mu_C)
(n_\pi , 0); \rho (\lambda ,\mu ) \kappa LM\rangle
~~~,
\label{eq-1b}
\eeqa
where $\rho_C$, $\rho$ and $\kappa$ are multiplicity labels.
The cluster irreps are coupled first to $(\lambda_C,\mu_C)$,
then with $(n_\pi 0)$ to the final $SU(3)$ irrep
$(\lambda , \mu )$.
The advantage of using the ket-formalism is the absence
to the need of a coordinate space description, it suffices
to get the quantum numbers describing Pauli allowed cluster
states. Of course, the disadvantage is that no explicit 
space distribution of the clusters are depicted.

This representation of the model space is of advantage,
because it involves only $SU(3)$ groups, which refer
to the shell model space of the clusters and the complete
nucleus.

The word {\it Semi}  in the name of SACM appears due to the phenomenological character
of the Hamiltonian, which is a sum of terms 
associated to the single particle energy, 
quadrupole-quadrupole interactions, angular momentum operators, etc. Note, that this is an additional
ingredient and the construction of the cluster space
is independent of it, which can be used in any microscopic 
model.

Finally, we expose the path on how to deduce the 
$\widetilde{SU}(3)$ shell model space: Each shell
$\eta$ has $\frac{1}{2}(\eta +1)(\eta +2)$
orbital degrees of freedom. Taking into account the two
spin degrees of freedom, the group-chain classifying
the states within the shell $\eta$ is given by

\beqa
U((\eta +1 )(\eta + 2)) & \supset~~~
U(\frac{1}{2}(\eta + 1)(\eta + 2)) \otimes U_S(2)
\nonumber \\
\left[ 1^N\right] & ~~~~~~~~~~~ \widetilde{\left[ h\right]}
~~~~~~~~~~~~~~~~~~~ \left[ h\right]
\nonumber \\
U(\frac{1}{2}(\eta + 1)(\eta + 2)) & \supset ~~~SU(3)
\nonumber \\
\left[ h\right] & ~~~~~~~~~~~ (\lambda , \mu )
~~~,
\label{class}
\eeqa
where $[h]$ is a short hand notation for 
$\left[h_1,h_2\right]$
and the tilde denotes the conjugate Young diagram where
rows and columns are interchanged. (\ref{class})
gives the relation of the spin-part (denoted by the index $S$)
and the orbital part, such that the complete state is
anti-symmetric ($\left[ 1^N\right]$, with $N$ as the
number nucleons in the shell $\eta$). 
In the case of light nuclei, instead of the spin group
$SU_S(2)$ the spin-isospin group $SU_{ST}(4)$ appears.
In contrast, in heavy nuclei the protons and neutron
have to be considered within different shells, which is the
reason why (\ref{class}) is used.
For the reduction to
$SU(3)$ programs are available
\cite{bahri}.

The reduction scheme (\ref{class}) is applied to each shell,
which contains nucleons. Each 
$\Delta n_{\pi} \hbar\omega$
excitations ($\Delta n_{\pi} =0,1,...$) corresponds
to a particular
distribution of the nucleons within the shells.
The $SU(3)$ content
of each shell is multiplied with all others, 
resulting in a preliminary list.
Finally, the center of mass motion
is removed from the $\Delta n_\pi \hbar\omega$ excitation,
by multiplying the result for $0\hbar\omega$ with
$(\Delta n_\pi ,0)$, the $1\hbar\omega$ result by
$(\Delta n_\pi -1, 0)$, etc. and subtracting all the results 
from the before obtained large list. This is a standard 
procedure, which for heavy nuclei 
is applied separately to the proton and neutron part.

\subsection{Concepts for the extension to heavy nuclei}

When dealing with heavy nuclei one encounters 
an exploding number of states ($SU(3)$ irreps).
Thus, one not only has to find a consistent path
on how to combine the cluster irreps with the relative motion
to the final irreps, but also a way to restrict further
the irreps according to their importance to contribute at low energy. This has to be done twice, namely for protons and
neutrons, and combining them leads to even more irreps.
Thus, the method to restrict the Hilbert space is very essential.

For the extension of the SACM to heavy nuclei requires explanations of some facts, 
concepts 
and assumptions:

\begin{itemize}

\item All nucleons move in the {\it same mean field} of the parent
nucleus.
This is obvious, because even when clusters are defined as consisting of a subset of
$A_k$ ($k=1,2$) nucleons, these nucleons are still part of the parent nucleus 
and not free clusters. {\it All nucleons are part of the same 
shell model}, characterized by the same  oscillator energy
$\hbar\omega$ of the united nucleus.
As a consequence, a light cluster,
cannot be the same within the parent nucleus, as it is as a 
free nucleus. For example,
for a free and independent cluster the
$\hbar\omega$ value is already different. The identification
of the clusters is only via their content in their number of
protons and neutrons. 

In addition, in $\widetilde{SU}(3)$ each cluster has
nucleons in the normal and unique orbitals and only the ones
in the normal orbitals are counted. Thus, a cluster changes
to a {\it pseudo-cluster} with $\widetilde{A}_k$ 
nucleons and $\widetilde{Z}_k$ protons 
in the normal orbitals ($k=1,2$). 

\item Because protons and neutrons are in different shells, the 
construction of the model space is applied separately to them,
as done {\it in any shell model application for heavy nuclei}
\cite{ring}. The proton and neutron part move in the
same mean field, i.e., with the same $\hbar\omega$.

\item The nucleons are filled into the Nilsson 
orbitals {\it at the deformation of the united nucleus}. 
The deformation values for a
nucleus is retrieved from the tables \cite{nix-tables}.
This does not suggest that we are working in different shell models, it is just a fact that protons and neutrons are filling
different kind of
orbitals, which is obvious by inspecting the Nilsson
diagrams for protons and neutrons \cite{ring}. 
For both, however, the
deformation value {\it is the same}.

\item The nucleons of the heavy cluster are filled in first and then the ones of the light clusters.
This involves the assumption that the light cluster is small,
compared to the heavier one, 
and is preformed within the united nucleus, a 
phenomenological picture used in radioactive decays.

This step will provide us with a heavy and light cluster, for each one with a certain number of nucleons in the normal 
orbitals, denoted by $\widetilde{A}_k$ ($k=1,2$).

\item In a final step, the proton and neutron part are 
combined, proposing a particular selection, whose
origin is the  demand that the proton and neutron fluid
are aligned. In \cite{draayer1} this is discussed
and in more detail in \cite{NPA576}: In the nuclear
shell model the aligned irrep, i.e.
$(\lambda_p ,\mu_p ) \otimes 
(\lambda_n ,\mu_n )$ $\rightarrow$
$(\lambda_p + \lambda_n , \mu_p + \mu_n )$
($p$ for protons and $n$ for neutrons), 
corresponds to
aligned principle axes of the two (proton-neutron) rotors.
All other irreps are non-aligned proton-neutron
rotors, which lie at higher energy, corresponding to
scissors mode, or isovector resonances \cite{eisenberg}.
Thus, the restriction to aligned proton-neutron irreps
is a pretty good approximation, taking effectively into
account the interaction, which
tends to align the proton and neutron fluid.
 
\end{itemize}
\vskip 0.5cm

\subsection{Construction of the model space and further restrictions}

The construction of the ${\widetilde SU}(3)$
model space for heavy nuclei is
in line with the one for light nuclei.
First some notational definitions: An $\alpha =p,n$ denotes the type of the nucleons (protons or neutrons), 
$(\widetilde{\lambda}_{k \alpha},\widetilde{\mu}_{k \alpha})$ ($k = 1,2$) refers
to cluster number $k$ and for each type of nucleon. The
$(\widetilde{\lambda}_{\alpha C},
\widetilde{\mu}_{\alpha_C})$ is
called the {\it cluster irrep} 
for the $\alpha$-type nucleons and it is the
result of the coupling

\beqa
(\widetilde{\lambda}_{1 \alpha},\widetilde{\mu}_{1 \alpha}) \otimes
(\widetilde{\lambda}_{2 \alpha},\widetilde{\mu}_{2 \alpha})
& = & \sum_{\widetilde{\lambda}_{\alpha C},
\widetilde{\mu}_{\alpha_C}}
m_{\widetilde{\lambda}_{\alpha C},
\widetilde{\mu}_{\alpha_C}}
(\widetilde{\lambda}_{\alpha C},
\widetilde{\mu}_{\alpha_C})
~~~,
\label{coupl1}
\eeqa
being $m_{\widetilde{\lambda}_{\alpha C},\widetilde{\mu}_{\alpha_C}}$ the multiplicity
of $(\widetilde{\lambda}_{\alpha C},
\widetilde{\mu}_{\alpha_C})$.

Afterward, each $(\widetilde{\lambda}_{\alpha C},\widetilde{\mu}_{\alpha_C})$
is coupled with the relative motion irrep 
$(\widetilde{n}_{\alpha\pi} , 0)$, i.e.,

\beqa
(\widetilde{\lambda}_{\alpha C},\widetilde{\mu}_{\alpha_C}) \otimes
(\widetilde{n}_{\alpha\pi} , 0) & = &
\sum_{\widetilde{\lambda}_\alpha , \widetilde{\mu}_\alpha }
m_{\widetilde{\lambda}_\alpha , \widetilde{\mu}_\alpha} 
(\widetilde{\lambda}_\alpha , \widetilde{\mu}_\alpha )
~~~.
\eeqa
The minimal number of
${\widetilde n}_{\alpha\pi}$ is determined in the
same manner as in the SACM for light nuclei: It is the
difference of the number of $\pi$-oscillation 
quanta in the
united nucleus (only counting those in the normal orbitals)
and the sum of oscillation quanta of the two clusters,
for protons and neutrons separately.

This leads to a large list of irreps
$(\widetilde{\lambda}_\alpha , \widetilde{\mu}_\alpha )$ for each sector of nucleons.
This list is compared to the one
obtained by the shell model. A program, which does this automatically, can be send on request.

For a two cluster system,
the path explained is resumed in the following group chains:

\beqa
&
\widetilde{SU}_{1\alpha}(3)
\otimes
\widetilde{SU}_{2\alpha}(3)
\otimes 
\widetilde{SU}_{R\alpha}(3) \supset
\nonumber \\
&
(\lambda_{1\alpha},\mu_{1\alpha}) ~~
(\lambda_{2\alpha},\mu_{1\alpha}) ~~
(\widetilde{n}_{\alpha\pi},0) 
\nonumber \\
&
\widetilde{SU}_{\alpha C}(3) \otimes 
\widetilde{SU}_{R\alpha}(3) \supset \widetilde{SU}(3) 
\nonumber \\
&
~~(\lambda_{\alpha C},\mu_{\alpha C})~~~
(\widetilde{n}_{\alpha\pi},0)~~~ 
~~~(\widetilde{\lambda},\widetilde{\mu})
~~~,
\eeqa
where $R$ refers to the relative motion part. Below each group
the corresponding quantum numbers are listed. The
$\widetilde{SU}(3)$ can be further reduced to the 
angular momentum group.

Up to here, the procedure is in accordance to the SACM for
light nuclei \cite{cseh-letter}, with the difference that due to the
breaking of isospin symmetry the protons and neutrons are treated separately.
In order to obtain the final list of Pauli allowed
total irreps, the one of protons is multiplied with the one of 
neutrons, i.e.,

\beqa
(\widetilde{\lambda}_{p},\widetilde{\mu}_{p}) 
\otimes (\widetilde{\lambda}_{n},\widetilde{\mu}_{n})
& = &
\sum_{\widetilde{\lambda} , \widetilde{\mu}} 
m_{\widetilde{\lambda} , \widetilde{\mu}} 
(\widetilde{\lambda} , \widetilde{\mu} )
~~~,
\label{large}
\eeqa 
which is represented by the following group chain

\beqa
\widetilde{SU}_p (3) \otimes \widetilde{SU}_n (3) 
& \supset & 
\widetilde{SU}(3)
~~~.
\eeqa
To reduce further the large list, obtained 
in (\ref{large}), we select only the 
irreps corresponding to a linear coupling,
namely

\beqa
(\lambda_p,\mu_p) \otimes (\lambda_n,\mu_n)
& \rightarrow &
(\lambda_p + \lambda_n,\mu_p + \mu_n)
~~~.
\label{linear}
\eeqa
The justification is the same as mentioned further above,
i.e., all other irreps correspond to scissors 
modes at large high energy \cite{draayer1}.

\subsection{The concept of forbiddenness}

As for large clusters in light nuclei, for heavy clusters
in heavy nuclei an additional problem arises:

Usually, the clusters are in their ground state and
all shell excitations are dealt via the radial excitation.
This is because an excitation of the clusters is equal
to the excitation in the radial motion. 

However, as mentioned above, for the light cluster being 
sufficiently large the clusters have to be excited, 
in order to connect to the ground state 
irrep of the parent nucleus. The 
required number of excitation quanta are 
subtracted from the relative motion. 
The cluster system has to be excited
by a minimal number of shell excitation $n_C$ =
$n_{pC}+n_{nC}$, such that for the first time 
the ground state can be reached in the coupling of the 
cluster irreps and the relative motion,
containing now less number of quanta. Further excitations
have to be added to the relative motion with the same argument as given in the last paragraph.
In \cite{smirnov} this concept was denoted {\it forbiddenenness}
with important consequences for the explanation 
of the preferences
of fission. This concept is barely known but applied with success
within the SACM 
\cite{sacm-fission1,sacm-fission2,sacm-fission3}. 
In the examples presented in section \ref{examples} 
we will demonstrate the importance of the
{\it forbiddenness}.

An easy to apply formula to determine the 
{\it forbiddenness} $n_{\alpha C}$
is given in \cite{huitz-2015}
(again for protons and neutrons separately, with 
$\alpha =p$ or $n$, and
all variables are denoted by a tilde in order to stress that we
work within the $\widetilde{SU}(3)$ language):

\beqa
&
{\widetilde n}_{\alpha C}  = 
&
\nonumber \\
& 
{\rm max}\left[0,\frac{1}{3}\left\{ {\widetilde n}_{\alpha 0}-
({\widetilde \lambda}_\alpha -{\widetilde \mu}_\alpha
)-(2{\widetilde \lambda}_{\alpha C}+{\widetilde \mu} _{\alpha C})\right\} \right] 
&
\nonumber \\
& 
+{\rm max}\left[ 0, \frac{1}{3}\left\{ {\widetilde n}_{\alpha 0}-
({\widetilde \lambda}_\alpha +2{\widetilde \mu}_\alpha )
+({\widetilde \lambda} _{\alpha C}-{\widetilde \mu}_{\alpha C})\right\} \right]
~~~,
&
\nonumber \\
&&
\label{forbid-min}
\eeqa
where $\widetilde{n}_{\alpha 0}$ is the minimal number
of relative oscillation quanta for $\alpha$ (p or n)
as required by the Wildermuth
condition. The {\it forbiddenness} can 
be zero, as is the case for most cluster systems of light nuclei.
The $({\widetilde \lambda}_{\alpha C},{\widetilde \mu}_{\alpha C})$ denotes 
the cluster irrep (to which the two clusters
are coupled and there may appear several)
and $({\widetilde \lambda}_\alpha , {\widetilde \mu}_\alpha )$ 
is the final $SU(3)$ irrep of the
united nucleus.
For later use, we define 
$\left({\widetilde \lambda}_{\alpha 0}, {\widetilde \mu}_{\alpha 0}\right)$ 
as the difference of the excited cluster
irrep $\left({\widetilde \lambda}^e_{\alpha C},{\widetilde \mu}^e_{\alpha C}\right)$ 
to the non-excited one 
$\left({\widetilde \lambda}_{\alpha C},{\widetilde \mu}_{\alpha C}\right)$ 
(the letter $e$ stands for {\it excited}), via

\beqa
\left({\widetilde \lambda}^e_{\alpha C},
{\widetilde \mu}^e_{\alpha C}\right) & = & 
\left({\widetilde \lambda}_{\alpha C} + 
{\widetilde \lambda}_{\alpha 0},
{\widetilde \mu}_{\alpha C}+{\widetilde \mu}_{\alpha 0}\right)
~~~.
\label{la0mu0}
\eeqa

Eq. (\ref{forbid-min}) 
can be interpreted as follows: 
The first term in (\ref{forbid-min}) indicates, that 
in order to {\it minimize} ${\widetilde n}_{\alpha C}$, we
have to {\it maximize} $({\widetilde \lambda}_{\alpha C}+2{\widetilde \mu}_{\alpha C})$.
The second term suggests to minimize the difference
$\left({\widetilde \lambda}_{\alpha C} - 
{\widetilde \mu}_{\alpha C}\right)$. 
The condition of a maximal 
$({\widetilde \lambda}_{\alpha C}+
2{\widetilde \mu}_{\alpha C})$ and a minimal 
$\left({\widetilde \lambda}_{\alpha C} - 
{\widetilde \mu}_{\alpha C}\right)$
implies a large compact and oblate 
configuration of the two-cluster system.
Information on
the relative orientation of the clusters is obtained in comparing the distribution
of oscillation quanta for each cluster in the different spatial directions 
\cite{NPA576,orient}.

These conditions are achieved, 
determining the whole product
of $({\widetilde \lambda}_{1\alpha},{\widetilde \mu}_{1\alpha}) 
\otimes ({\widetilde \lambda}_{2\alpha},
{\widetilde \mu}_{2\alpha})$, using (\ref{coupl1})
and searching for the irrep that corresponds to a large
compact structure. The
${\widetilde n}_0={\widetilde n}_{p0}+{\widetilde n}_{n0}$ 
is the total minimal number of relative
excitation quanta and 
${\widetilde n}_C={\widetilde n}_{pC}+{\widetilde n}_{nC}$ is the total
{\it forbiddenness}.

The $\widetilde{n}_{\alpha C}$ are transferred to the
clusters, where the result does not depend
on how the distribution is done.
For example, when only one cluster is excited, 
first the irrep of the number of one less valence nucleons
is determined
and then coupled with the 
nucleon in the higher shell, with $n_{\alpha C}$ quanta
above the valence shell. 
The important criterion
is that at the end one finds a combination of irreps
which couple to the final irrep in the united cluster.
This is a direct but rather cumbersome procedure and
how to do it will be illustrated in the applications. 

More restrictions have to be implemented, in order to avoid
a too large, not manageable list of irreps:

\begin{itemize}

\item When coupling the cluster irreps, only those
$\left( \lambda_{\alpha C},\mu_{\alpha C}\right)$
irreps in the product are considered which couple to the
ground state of the united nucleus, for protons and neutron
separately.

\item Of those, only the ones are considered with the largest
eigenvalues of the second order Casimir operator
of ${\widetilde SU}(3)$, often the irrep with the largest
eigenvalue suffices. The main argument is that in deformed
nuclei, with a significant quadrupole-quadrupole interaction,
only these selected irreps contribute 
significantly at low energy.

\item In coupling protons and neutrons together, only the
linear coupling is considered, for reasons explained before.

\item The final list is, in general, still too large and
one has to restrict to the first few irreps with the larges 
eigenvalues of the second order Casimir operator, an argument
valid for  dominant quadrupole-quadrupole interaction.

\item {\it Convergence criterium}:
Convergence due to these restrictions have to 
be verified by adding 
and/or restricting some irreps. If no sensible change
is observed, then the cut-off procedure is set.

\end{itemize}

This is a consistent procedure and 
provides a basis for the extension to heavy nuclei:
The sum of nucleons in the normal orbitals, of the two 
cluster,  will be always equal to the ones in the united 
nucleus, which is for example ignored in \cite{cseh-algora}.
All procedures known from SACM can now be applied in the same manner.

For completeness, we mention a different definition for the {\it forbiddenness}:
In order to be able to deal with heavy systems, 
the alternative definition was
proposed in \cite{cseh-algora,cseh-scheid}, 
where the relation of an irrep to its deformation 
was exploited \cite{rowe1}.
The criterion applied is as follows: 
If an irrep of the list has a similar deformation as the one in the shell model allowed irrep, it implies a {\it less forbidden} state than 
irreps with a larger difference in the irreps. 
The reason why it works is that the deformation can
be related to $SU(3)$ irreps \cite{rowe1,TE} and comparing
deformations is equivalent to comparing the
dimension of these irreps.

Therefore, the definition of {\it forbiddenness} used 
in \cite{cseh-scheid,cseh-algora} is 

\beqa
F & = & \frac{1}{1+min\left[\sqrt{\Delta n_1^2+\Delta n_2^2+\Delta n_3^2}\right]}
~~~,
\label{eq-3}
\eeqa
where $\Delta n_i = \mid n_i - n_{i,\xi}\mid$
and in contrast to $S$, as defined in  
\cite{cseh-scheid,cseh-algora}, we use $F$ because the letter
$S$ will be used later exclusively for the spectroscopic factor. 
The index $i$ refers to the 
spatial direction of the oscillation and 
$\xi$ to the  several cluster irreps allowed by  the Pauli-exclusion principle. The $n_i$ is the
number of oscillation quanta in direction $i$. 
When all $\Delta n_i$ are zero then $F=1$ and the irrep is allowed. If at least one of those numbers is different from
zero the irrep is partially forbidden and when $F=0$ 
it is completely forbidden.

\subsection{The structure of the Hamiltonian and
the quadrupole transition operator}
\label{hamiltonian}

The most general algebraic Hamiltonian has the  same structure as for light nuclei, save that the
operators (number operator, quadrupole operator, etc.) are substituted by their pseudo counter
parts. How this mapping is achieved, is explained in detail in 
\cite{draayer1}, where an operator ${\bd O}$ in $SU(3)$ 
is mapped to its counterpart ${\bd O}$
$\approx$ $\kappa {\bd {\widetilde O}}$ and $\kappa$ has a simple
approximation, namely 
$\kappa = \frac{\eta + \frac{3}{2}}{\eta + \frac{1}{2}}$,
close to 1. The factor can be assimilated into the
parameters of the SACM Hamiltonian.

We restrict to the $SU(3)$ limit, which turns out to 
be sufficient, due to the large deformation of the systems 
considered. This, however, will result in zero
$B(E2)$ transition rates between states belonging
to different ${\widetilde SU}(3)$ irreps.

A simplified
model Hamiltonian is selected, which
has the following structure:

\begin{eqnarray}
{\bd H} & = &
\hbar \omega \mbox{\boldmath$\widetilde{n}$}_{\pi }
+(a_2-a_5\Delta
\mbox{\boldmath$\widetilde{n}$}_{\pi })\mathit{\widetilde{{\bd C}}}_{2}
\left( \widetilde{\lambda} ,\widetilde{\mu} \right)  
\nonumber \\
&&
+ t_3 \left[ \mathit{\widetilde{{\bd C}}}_{2}
\left( \widetilde{\lambda} ,\widetilde{\mu} \right) \right]^2
+ t_1\mathit{\widetilde{{\bd C}}}_3
\left( \widetilde{\lambda} , \widetilde{\mu} \right)
\nonumber \\
&&
+\left( a_3 + a_{Lnp} \Delta \widetilde{{\bd n}}_\pi \right) 
{\mbox{\boldmath$\widetilde{L}$}}^{2}+t_2{\mbox{\boldmath$\widetilde{K}$}}^{2}
\label{ham}
\end{eqnarray}
where $\Delta \widetilde{{\bd n}}_\pi = 
\widetilde{{\bd n}}_\pi - ({\widetilde n}_0-{\widetilde n}_C)$, 
$({\widetilde n}_0-{\widetilde n}_C)$
being the total minimal number of quanta required by the Pauli principle
and the possible effects of the {\it forbiddeness} are taken 
into account by ${\widetilde n}_C$.
The moment of inertia, contained in the factor of
the total angular momentum operator, 
$\widetilde{{\bd L}}^2$,  
may depend on the excitation in $\tilde{n}_\pi$ 
(excited states change their deformation, corresponding to a variable momentum
of inertia). 
The ${\bd K}^2$ term serves to split the degeneracy in ${\widetilde {\bd L}}$
within the same $\widetilde{SU}(3)$ irrep.
The first term of the $\widetilde{SU}(3)$ Hamiltonian, $\hbar\omega \mbox{\boldmath
$\widetilde{n}$}_\pi$, contains the linear invariant operator of the  $\widetilde{U}_R(3)$ subgroup defining the mean field, 
and the $\hbar\omega$ is fixed via $(45 A^{-1/3}
- 25 A^{-2/3})$~MeV for
light nuclei \cite{hw}, which can also be used for heavy nuclei.
For heavy nuclei $\hbar\omega = 41 A^{-\frac{1}{3}}$~MeV is more common. The $A$ is the mass number of the real nucleus
and not the number ${\tilde A}=\widetilde{A}_1+\widetilde{A}_2$ 
of nucleons in the normal
orbitals for the united nucleus.

The $\widetilde{{\bd C}}_2\left({\widetilde \lambda},{\widetilde \mu}\right)$
is the second order Casimir-invariant of the coupled
$\widetilde{SU}(3)$ group, having contributions both from the internal
cluster part and from the relative motion. 
The $\widetilde{{\bd C}}_2\left({\widetilde \lambda},{\widetilde \mu}\right)$ is given by:

\begin{eqnarray}
\mbox{\boldmath $\widetilde{C}$}_2(\widetilde{\lambda},\widetilde{\mu}) & = & 2 \mbox{\boldmath $\widetilde{Q}$}^2 +
\frac{3}{4} \mbox{\boldmath $\widetilde{L}$}^2 ,  \nonumber \\
& \rightarrow & \left(\widetilde{\lambda}^2 + \widetilde{\lambda}\widetilde{\mu} 
+ \widetilde{\mu}^2 + 3\widetilde{\lambda} + 3\widetilde{\mu}
\right) ,  \nonumber \\
\mbox{\boldmath $\widetilde{Q}$} & = & \mbox{\boldmath $\widetilde{Q}$}_C 
+ \mbox{\boldmath $\widetilde{Q}$}_R ,
\nonumber \\
\mbox{\boldmath $\widetilde{L}$} & = & \mbox{\boldmath $\widetilde{L}$}_C 
+ \mbox{\boldmath $\widetilde{L}$}_R ,
\label{su3}
\end{eqnarray}
where $\mbox{\boldmath $\widetilde{Q}$}$ and 
$\mbox{\boldmath $\widetilde{L}$}$ are
the total quadrupole and angular momentum operators,
respectively, and $R$ refers to the relative motion. 
The eigenvalue of $\mbox{\boldmath $\widetilde{C}$}_2(\widetilde{\lambda},\widetilde{\mu})$ is also indicated.
The relations of the quadrupole and angular momentum
operators to the $\widetilde{C}^{(1,1)}_{2m}$ generators of the $\widetilde{SU}(3)$ 
group, expressed in terms of $\widetilde{SU}(3)$-coupled $\pi$-boson creation and
annihilation operators, are \cite{escher}:

\begin{eqnarray}
\mbox{\boldmath $\widetilde{Q}$}_{k,2m} & = & \frac{1}{\sqrt{3}}
\widetilde{C}^{(1,1)}_{k2m} , \nonumber \\
\mbox{\boldmath $\widetilde{L}$}_{k1m} & = & \widetilde{C}^{(1,1)}_{k1m} , 
\nonumber \\
\mbox{\boldmath $\widetilde{C}$}^{(1,1)}_{lm} & = & \sqrt{2} \left[
\mbox{\boldmath
$\pi$}^\dagger \otimes \mbox{\boldmath $\pi$} \right]^{(1,1)}_{lm} .
\label{su3gen}
\end{eqnarray}

The quadrupole electromagnetic transition operator is defined
as

\beqa
\bd{T}_m^{(E2)} & = & \sum_\gamma e_\gamma^{(2)}
\bd{Q}^{(2)}_{\gamma ,m}
~~~,
\label{trans-p}
\eeqa
where $e_\gamma^{(2)}$ is the effective charge of the
contribution to the quadrupole operator, coming from
the cluster $\gamma$ = $C_1$, $C_2$ and from the relative
motion $R$. The effective charges are determined as
explained in great detail in \cite{fraser-2012}.  

\subsection{Spectroscopic factors}
\label{specfac}

A successful parametrization of the 
spectroscopic factor, within the SACM for light nuclei, is given in 
\cite{specfac-draayer}:

\beqa
&
S = 
& 
\nonumber \\
&
e^{{\cal A} + B n_\pi + C{\cal C}_2(\lambda_1,\mu_1) 
+ D{\cal C}_2(\lambda_2,\mu_2) 
+ E{\cal C}_2(\lambda_c,\mu_c)}
&
\nonumber \\
&
\times
e^{F{\cal C}_2(\lambda , \mu ) + G{\cal C}_3(\lambda , \mu ) 
+ H\Delta n_\pi}
&
\nonumber \\
&
\mid 
\langle (\lambda_1,\mu_1)\kappa_1L_1, 
(\lambda_2,\mu_2)\kappa_2L_2 \mid\mid
(\lambda_C,\mu_C)\kappa_C L_C \rangle_{\varrho_C}
&
\nonumber \\
& \cdot
\langle (\lambda_C,\mu_C)\kappa_CL_C, 
(n_\pi ,0)1l \mid\mid
(\lambda ,\mu )\kappa L \rangle_1
\mid^2
~~~,
&
\label{specfac-light}
\eeqa
where the $\varrho$-numbers refer to multiplicities in the coupling to $SU(3)$ irreps and the $\kappa$'s to the 
multiplicities to the reduction to $SO(3)$.
The parameters were adjusted to theoretically exactly calculated 
spectroscopic factors within the p- and sd-shell,
using
the $SU(3)$ shell model \cite{draayer2}, with an excellent 
coincidence. For the good
agreement, the factor depending on the $SU(3)$-isoscalar factors turns out to be crucial.

For heavy nuclei, spectroscopic factors are poorly or not at all known experimentally.
Therefore, we have to propose a simplified manageable ansatz, compared
to (\ref{specfac-light}), including the {\it forbiddenness}
and, if possible, {\it parameter free}.

In what follows, we will propose an expression for the
spectroscopic factor which is motivated by the one
used for nuclei in the p- and sd-shell. 
As in (\ref{specfac-light}) the expression
is divided into two factors, 
the first one is an exponential factor and the
second one of pure geometrical origin \cite{specfac-draayer}, 
an expression in terms
of coupling coefficients. The second factor is maintained,
because it refers to coupling of $SU(3)$ irreps only.  

The first exponential factor deserves more explanation:
As argued in \cite{specfac-draayer}, 
this term is the result of the relative part of the wave-function, which 
for zero angular momentum is proportional to 
$e^{-aR^2} \sim e^{-a\frac{\hbar}{\mu \omega} n_\pi}$, where
$R$ is the relative distance of the two clusters (though, an $e^{-a R}$ ansatz 
would be more appropriate, but the clusters are defined in the harmonic oscillator picture
and we stay to it for consistency)
and $a$ has units of $\rm{fm}^{-2}$. 
The $\mu$ is the reduced mass. Let us
restrict to the minimum value $n_0$ of $n_\pi$. Using the relation of 
$r_0 = \sqrt{\frac{\hbar}{\mu\omega}n_0}$ \cite{geom}, where $r_0$ is
the average, minimal distance between the clusters, and taking
into account that for this case $R=r_0$, we obtain
$e^{-\mid B\mid n_0}$, with 
$\mid B\mid = a\frac{\hbar}{\mu\omega}$
and $B<0$.
When the wave function acquires the value $e^{-1}$ it results in the relation 
$\mid B\mid =\frac{1}{n_0}$. 
For the nuclei in the sd-shell, the 
adjustment of the parameters was done for cases with $n_0=8$,
which corresponds according to this estimation to 
$B \approx -0.13$. 
This has to be compared to the value  $-0.36$ as obtained in
the fit in 
\cite{specfac-draayer}, i.e., it is only an approximation which gives at least the
correct order. The most
important part of (\ref{specfac-light}) is the factor
depending on the $SU(3)$ isoscalar factors,
which are responsible for the relative changes, while
the influence of the exponential factor is not dominant for the relative numerical values of the
spectroscopic factors.
Furthermore, when {\it ratios} of spectroscopic factors are used,
the exponential contribution cancels for states in the $0\hbar\omega$
shell. Also when the $(n_0-nc)$ is large, the corrections
for $\Delta n_\pi$ of the order of one will be negligible.
We do not see any possibility to estimate the parameter 
$\cal{A}$ in
the exponential factor, which represents a normalization.
The other terms in the exponential factor represent corrections
to the inter-cluster distance, because they correspond
to deformation effects and the parameters in front turned out
to be consistently small, thus, they can for the
moment be neglected. 

In light of the above estimation and discussion, 
for heavy nuclei we propose a similar expression as
in (\ref{specfac-light}), but due to the non-availability 
of a sufficient number of spectroscopic factor values (or none at all)
for heavy nuclei, we propose the following simplified expression:

\beqa
&
S = 
& 
\nonumber \\
&
e^{{\tilde {\cal A}} 
+ {\tilde B} ({\tilde n}_0-{\tilde n}_C+\Delta {\tilde n}_\pi )}
&
\nonumber \\
&
\mid 
\langle ({\tilde \lambda}_1,{\tilde \mu}_1)
{\tilde \kappa}_1{\tilde L}_1, 
({\tilde \lambda}_2,{\tilde \mu}_2)
{\tilde \kappa}_2{\tilde L}_2 \mid\mid
({\tilde \lambda}_C+{\widetilde \lambda}_0,{\tilde \mu}_C
+{\widetilde \mu}_0){\tilde \kappa}_C
{\tilde L}_C \rangle_{\varrho_C}
&
\nonumber \\
& \cdot
\langle ({\tilde \lambda}_C+{\widetilde \lambda}_0,
{\tilde \mu}_C+{\widetilde \mu}_0)
{\tilde \kappa}_C{\tilde L}_C, 
({\tilde n}_\pi ,0)1{\tilde l} \mid\mid
({\tilde \lambda} ,{\tilde \mu} ){\tilde \kappa} 
{\tilde L} \rangle_1
\mid^2
~~~.
&
\label{specfac-heavy}
\eeqa 
The $n_\pi$ in the exponential factor was substituted
by $[(\tilde{n}_0-\tilde{n}_C)+\Delta n_\pi ]$. The $({\tilde n}_0-{\tilde n}_C)$ is the
number of relative oscillation quanta in $0\hbar\omega$ ($\tilde{n}_C$ was added to
the excitation of the clusters).

The parameter is estimated as 
${\tilde B}=-\frac{1}{{(\tilde n}_0-\tilde{n}_C)}$. Because we can not determine
the parameter ${\tilde {\cal A}}$, as a consequence 
only spectroscopic factors divided by the exponential factor $e^{{\widetilde A}}$ are listed.
An additional dependence on $\tilde{n}_C$ is contained in the product of 
reduced coupling coefficients, with the appearance of 
${\widetilde \lambda}_0$
and ${\widetilde \mu}_0$ (see the definition in (\ref{la0mu0})). 
${\widetilde n}_0$ and ${\widetilde n}_C$ refer to the values within the
parent nucleus. In general, in (\ref{specfac-heavy}) further
parameters can be added, as for light nuclei, which have
though to be adjusted.

With this choice, the spectroscopic factor is 
{\it parameter free},
except for an overall normalization, which does not
play a role when only ratios are of interest.

\section{Applications}
\label{examples}

In this section we apply the pseudo-SACM to three
sample systems. The first is
$^{236}_{92}$U$_{144}$ $ \rightarrow$ 
$^{210}_{82}$Pb$_{128}$+$^{26}_{10}$Ne$_{16}$, 
the second one is
$^{224}_{88}$Ra$_{136}$ $\rightarrow$ 
$^{210}_{82}$Pb$_{128}$+$^{14}_{6}$C$_{8}$ 
and the third one is $^{236}_{92}$U$_{144}$ $\rightarrow$ 
$^{146}_{54}$Xe$_{92}$+$^{90}_{38}$Sr$_{52}$,
which appears in the fission channel of $^{236}$U.
In the first two cases, the {\it forbiddeness} 
is small. They serve to illustrate how to add the quanta to the clusters and how to determine the final cluster irrep.
The importance of the {\it forbiddeness} is significant in
the third case and it will be much harder to determine the cluster irreps. This sequence also shows that the larger
the lighter cluster is, the larger the {\it forbiddenness} 
becomes.

For illustrative reasons, only the $\widetilde{SU}(3)$ dynamical symmetry limit is
considered, i.e., the united nucleus must be well deformed.
The mixing of $SU(3)$ irreps will be investigated in a
future publication.
The examples also serve to illustrate the 
applicability of the model and on
how to deduce the quantum numbers. However, 
some transitions will
be forbidden due to $SU(3)$ selection rules.

Very useful is the Table \ref{tabnum}, where the number
of nucleons in a given shell $\eta$ is related to the
one of the pseudo-shell number $\widetilde{\eta}$.
In the last columns the accumulated number of oscillation
quanta within the $\widetilde{SU}(3)$ shell model
is listed.

\begin{center}
\begin{table}[h!]
\centering
\begin{tabular}{|c|c|c|c|c|}
\hline\hline
$\eta$ & No. of nucleons & $\widetilde{\eta}$ & No. of nucleons & acum. No. of quanta\\
\hline
0 & 2 & - & - & - \\
1 & 6 & 0 & 2 & 0 \\
2 & 12 & 1 & 6 & 6 \\
3 & 20 & 2 & 12 & 30 \\
4 & 30 & 3 & 20 & 90 \\
5 & 42 & 4 & 30 & 210 \\
6 & 56 & 5 & 42 & 420 \\
\hline 
 \end{tabular}
\caption{
Table, listing the number of particles in the
shell number $\eta$ and the pseudo-shell number
$\widetilde{\eta}$, including the spin degree of freedom. 
The last column lists
the accumulated number of quanta of the $\widetilde{SU}(3)$, 
when each shell is
full up to the valence one. The final total number of quanta
is obtained, by adding to the number of quanta, reached up
to the last closed shell, the number of quanta of the
nucleons in the valence shell.
} 
\vspace{0.2cm}
\label{tabnum}
\end{table}
\end{center}

In this section, we explain trough the examples 
in great detail
the calculations which lead us to the cluster irreps and
with the relative motion to the ground state irreps in
the proton and neutron part. This requires a lot of 
cumbersome counting and determination of irreps,
especially when relative oscillation quanta from
the {\it forbiddenness} are added to the clusters.
We apologize for that and ask the reader to be patient.
If she (he) is not interested in the details, 
she (he) just can
skip this part and jump to the final numbers of the cluster
irreps.

\subsection{$^{236}_{92}$U$_{144}$ 
$\rightarrow$ $^{210}_{82}$Pb$_{128}$+$^{26}_{10}$Ne$_{16}$}
\label{3.1}

\begin{table}[H]
	\begin{center}
		\begin{tabular}{|c| c c c c c | c|c|}
		\hline
		$\widetilde{\eta}$&0&1&2&3&4&&$N$ \\
		\hline
		$^{128}_{46}Pd_{82}$&2&6&12&20&6&2(0)+6+2(12))&\\
                                                            &  &  &    &    &   &       +3(20)+4(6 & 114\\
		\hline
		$^{112}_{40}Zr_{72}$&2&6&12&20&0&2(0)+6+2(12)&\\
                                                           &   &  &   &    &   &      +3(20) & 90 \\
		\hline
		$^{14}_{6}C_{8}$&2&4&0&0&0&2(0)+4&4\\
		\hline
		\end{tabular}
	\caption{An example on how to determine the number
	of oscillation quanta ($N$) in the system derived from $^{236}$U
	$\rightarrow$ $^{210}$Pb+$^{26}$Ne, 
	with the help of Table 
	\ref{tabnum}. The same is applied for the two other
	sample cases, where Table \ref{tabnum} is very helpful
	for counting the number of relative excitation quanta.
	When one cluster is excited by $\widetilde{n}_C$ quanta,
	this number has still to be added.
	\label{wildpro}}
	\end{center}
\end{table}

\begin{table}[]
	\begin{center}
		\begin{tabular}{|c|c|c|c|}
		\hline
\textbf{Parameter} & system a & system b & system c  \\ 
\hline
$\hbar\omega$ & $6.63$ & $6.73$ & $6.63$  \\ \hline
$a_2$ & $-0.020841$ & $-0.0082194$ & $-0.023251$ \\ \hline
$a_3$ & $0.0087042$ & $0.0099524$ & $0.0067124$  \\ 
\hline
$t_1$ &  $6.698\times 10^{-5}$ & $-0.00061005$ & $-2.74\times 10^{-5}$  \\ \hline
$t_2$ & 0.40 & $0.2205$  & $0.0364275$  \\ 
\hline  		
$a_5$ & -0.0081127 & $0.1$ & $0.0098151$ \\ 
\hline  		
$a_L$ & $-0.0012576$ & $0.0025050$ & $0.00073054$ \\ 
\hline  		
$a_{L_{n_p}}$ & $6.55\times 10^{-3}$ & $0.00025837$ & $-0.00018069$ \\ 
\hline  		
$t_3$ & $-1.420\times 10^{-7}$ & $1.10\times 10^{-5}$ & $5.10\times 10^{-7}$ \\ 
\hline  		
\end{tabular}
	\caption{Non-zero parameter values for system a: 
	$^{236}$U
	$\rightarrow$ $^{210}$Pb + $^{26}$Ne;
	system b:
	$^{224}$Ra
	$\rightarrow$ $^{210}$Pb + $^{14}$C
	and system c: 
	$^{236}$U
	$\rightarrow$ $^{148}$Xe + $^{90}$Sr.}
	\label{224Ra-210Xe-parameters}
	\end{center}
\end{table}

\begin{figure}
\centerline{
\rotatebox{270}{\resizebox{200pt}{130pt}{\includegraphics[width=0.23\textwidth]{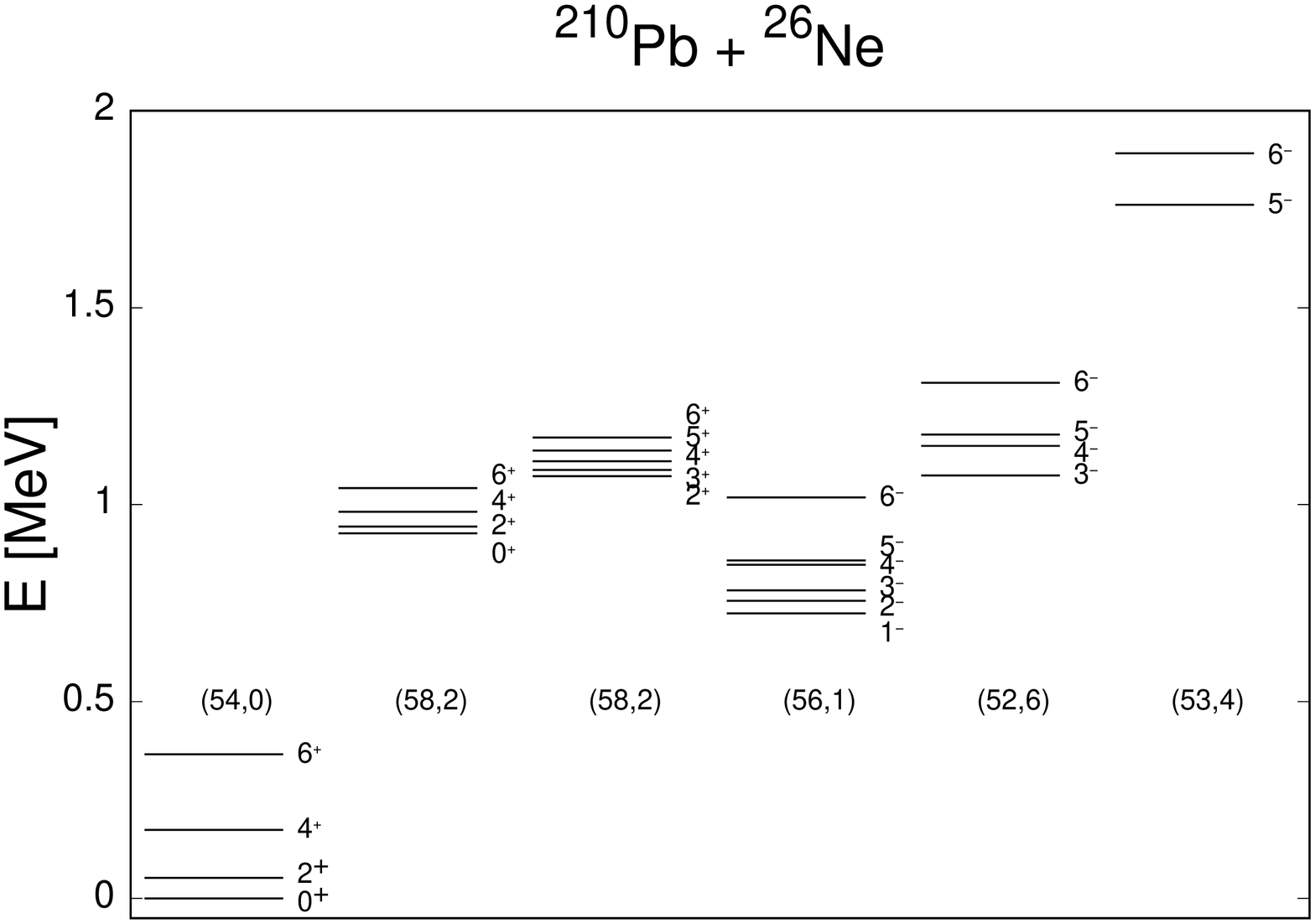}}}
\rotatebox{270}{\resizebox{200pt}{130pt}{\includegraphics[width=0.23\textwidth]{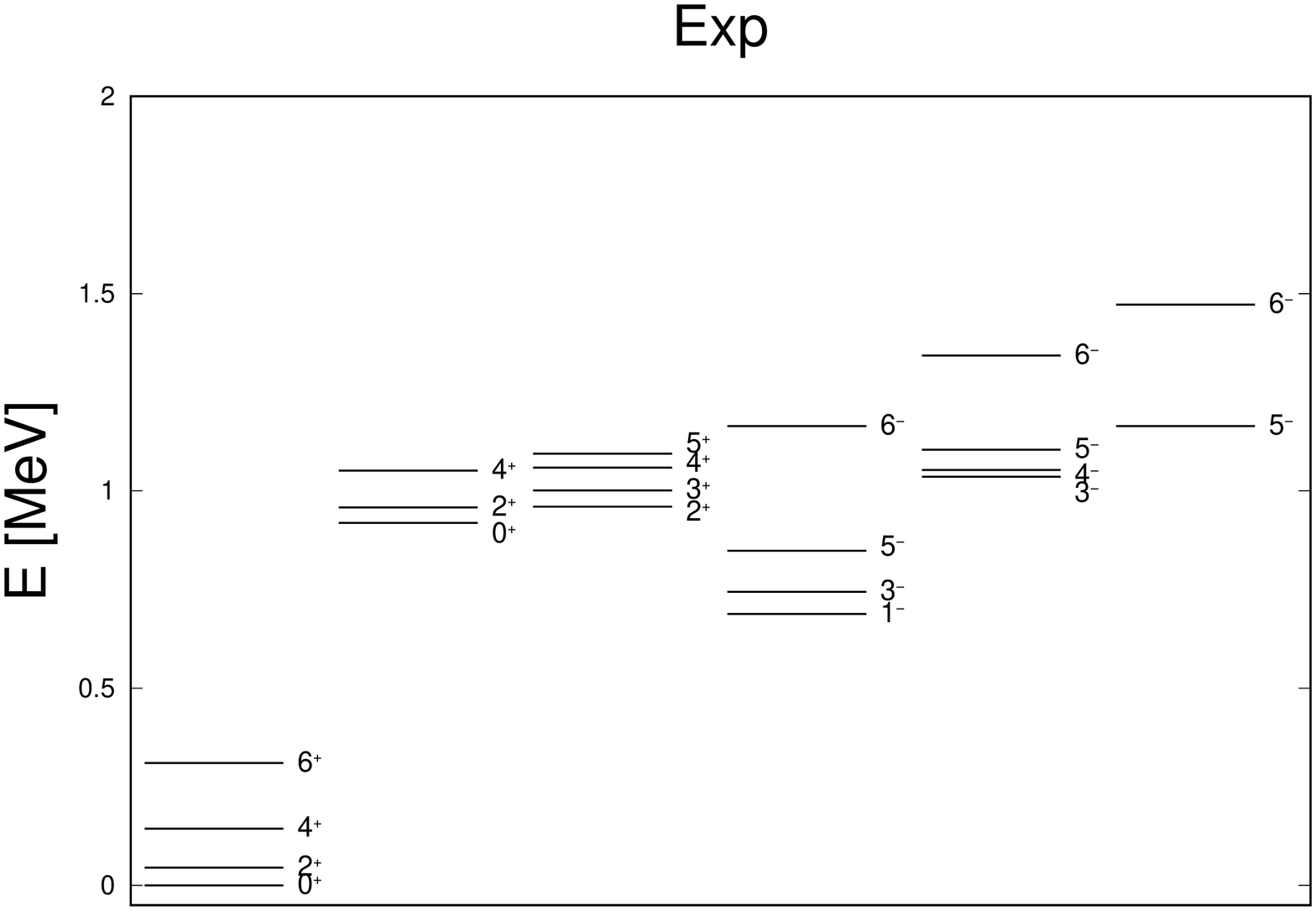}}}
}
\caption{\label{Uran} 
Spectrum of $^{236}$U, described by the 
clusterization $^{210}$Pb+$^{26}$Ne. 
Only states up to angular momentum 6 are depicted.
The theoretical spectrum
(left panel) is compared to experiment (right panel).
Below each rotational band the content of the
number of $\pi$ bosons ($n_{\pi}$) and the 
${\widetilde SU}(3)$ irrep is indicated.
}
\end{figure}

As explained above,
the protons and neutrons are treated separately and the nucleons in each sector
are filled into the Nilsson
diagram from below, at the 
deformation value $\epsilon_2 =0.200$ 
($\beta_2=0.215$) \cite{nix-tables}.
The $\hbar\omega = 6.63$~MeV.

For $^{236}$U, the united nucleus, we obtain 46 protons 
in the normal orbitals and the valence shell is
$\widetilde{\eta}_p=4$ with 6 valence protons. The ground state 
$\widetilde{SU}(3)$ irrep for the proton part
is $(\tilde{\lambda} , \tilde{\mu})_{p} =(18,0)_p$, while for the neutrons  there are  82 particles 
in normal orbitals with
12 in the $\widetilde{\eta}_n=5$ valence shell, giving the ground state irrep
$(\tilde{\lambda} , \tilde{\mu})_{n} =(36,0)_n$.
These two irreps can be coupled to the total one for $^{236}$U, namely 
$(\tilde{\lambda} , \tilde{\mu}) =(54,0)$.
The determination of the ground state irrep is necessary for the evaluation of 
the {\it forbiddenness} (see (\ref{forbid-min})). 

These considerations have to be repeated for the two clusters involved, with $^{210}$Pb being the largest
cluster and $^{14}$C the lightest one.
For $^{210}$Pb, filling 
the protons into the Nilsson diagram, at the same deformation as for the
united nucleus, we obtain 40 protons in normal orbitals, where the valence shell is
${\widetilde \eta}=3$ and closed, thus the corresponding irrep is $(0,0)^{\rm Pb}_{p}$. For the neutrons one has
72 in normal orbitals with 2 neutrons in the 
${\widetilde \eta}_n=5$ pseudo-shell. The corresponding
irrep is $(10,0)^{\rm Pb}_{n}$.

\begin{center}
\begin{table}[h!]
\centering
\begin{tabular}{|c|c|c|}
\hline\hline
$J_k^P$ & $^{236}$U $E_{{\rm exp}}$ [MeV] & $^{224}$Ra $E_{{\rm exp}}$ [MeV]  \\
\hline
$0_2^+$ & 0.919 & 0.916  \\
$2_1^+$ & 0.045 & 0.084  \\
$2_2^+$ & 0.958 & - \\
$2_3^+$ & 0.960 & 0.993 \\
$3_1^+$ & 1.002 & - \\
$4_1^+$ & 0.150 & 0.251 \\
$6_1^+$ & 0.210 & 0.479 \\
$1_1^-$ & 0.688 & 0.216 \\
$1_2^-$ & 0.967 & - \\
$3_1^-$ & 0.744 & 0.290 \\
$5_1^-$ & 0.848 & 0.433 \\
\hline
$J_i^P \rightarrow J_f^P$ & $^{236}$: $B(E2)$ [WU] & 
$^{224}$Ra: $B(E2)$ [WU] \\
\hline
$2_1 \rightarrow 0_1$ & 250. & 99. \\
$4_1 \rightarrow 2_1$ & 357. & 144 \\
\hline 
 \end{tabular}
\caption{
Experimental data used in the fit of the parameters of the model Hamiltonian. The second column lists
the data used for $^{236}$U 
and the third column for $^{224}$Ra.
If no data are mentioned (dashed sign), 
the experimental value is not used in the fit.
} 
\vspace{0.2cm}
\label{fit-U-Ra}
\end{table}
\end{center}

\begin{center}
\begin{table}[h!]
\centering
\begin{tabular}{|c|c|c|c|}
\hline\hline
$J_k^P$ & $^{236}$U-1 (th) & $^{224}$Ra (th)
& $^{236}$U-2 \\
\hline
$0_1^+$ & 0.0 & 0.0 & $0.00403$ \\
$0_2^+$ & 0.0 & 0.0 & $0.0000356$ \\
$2_1^+$ & 0.0 & 0.0 & $0.00384$ \\
$2_2^+$ & 0.0 & 0.0 & $4.79\times 10^{-7}$ \\
$4_1^+$ & 0.0 & 0.0 & $0.00346$ \\
$4_2^+$ & 0.0 & 0.0 & $7.00\times 10^{-6}$ \\
$1_1^-$ & 0.0 & 0.0 & $1.04\times 10^{-5}$ \\
$2_1^-$ & 0.0 & 0.0 & $1.04\times 10^{-5}$ \\
$3_1^-$ & 0.0 & 0.0 & $6.03\times 10^{-5}$ \\
\hline
\hline 
 \end{tabular}
\caption{
Some spectroscopic factors of low lying states, 
divided by $e^{\widetilde{A}}$. 
In the first column the state quantum numbers are tabulated.
The values of the spectroscopic factor for $^{236}$U 
containing the $^{210}$Pb cluster, $^{224}$Ra and
$^{236}$U containing the $^{146}$Xe cluster
are in the second, third and fourth column, respectively.
Only values which are greater than $10^{-8}$ are listed.
As seen, only the last system shows significant deviations
from zero.
} 
\vspace{0.2cm}
\label{Spec-U-Ra}
\end{table}
\end{center}

Using the numbers just deduced, the system
$^{236}_{92}$U $ \rightarrow$ 
$^{210}_{82}$Pb+$^{26}_{10}$Ne can be viewed
within the $\widetilde{SU}(3)$ description
as a $^{128}_{46}\widetilde{\rm{Pd}}$ 
$ \rightarrow$ 
$^{112}_{40}\widetilde{\rm{Zr}}$ +
$^{16}_{6}\widetilde{\rm{C}}$
cluster system, obtained by counting the protons and
neutrons in the normal orbitals.
Of course, these so-called {\it pseudo-nuclei} are
only schematic in nature.

The Tables \ref{tabnum} and \ref{wildpro} serve to illustrate
on how to determine the total number of quanta in each
cluster and the united nucleus, for the particular case
considered. For the other cases treated in this manuscript,
it will be similar.

The minimal number of quanta, which have to be added in the proton part, is 20 corresponding 
to a $(20,0)_{p R}$ 
irrep in the relative part. For the neutron part, this number is 40, i.e.,
an irrep $(40,0)_{n R}$.

In the next step, the proton part of the clusters are coupled with the relative part of the proton
section of the united nucleus. The same is done for the neutrons. For the proton part, the product
$(0,0)_p \otimes (0,2)_p \otimes (20,0)_{p R}$ 
(the index $pR$ refers to the relative motion of the
protons) contains the proton irrep $(18,0)_p$
of the united nucleus, thus, the {\it forbiddenness} for the proton part is zero. 
The situation is different for the neutron part: The product 
$(10,0)_n \otimes (4,0)_n \otimes (40,0)_{n R}$
{\it does not contain} (36,0), which is the irrep in the united nucleus.
This indicates that the clusters have to be excited, thus,
the {\it forbiddenness} is different from zero.
Using the formula (\ref{forbid-min}) we obtain a {\it forbiddenness} of $\widetilde{n}_C=2$. The excitation of the clusters
is achieved, changing as one possibility 
the irrep of $^{26}$Ne from $(4,0)_n$ to $(6,0)_n$. 
The relative part is now reduced by two 
quanta, leaving $(38,0)_{nR}$ 
(the index $nR$ refers to the relative motion of the
neutrons). With this change, the product
$(10,0)_n \otimes (6,0)_n \otimes (38,0)_{n R}$ 
now contains the dominant irrep for neutrons in $^{236}$U.
This is verified by Eq. (\ref{forbid-min}).
Distributing the excitation quanta in a different manner
leads to the same final result.

Using the Hamiltonian in the $SU(3)$-dynamical limit, the coefficients are adjusted to 
the experimental data, listed in Table \ref{fit-U-Ra} in the second column. 
The optimal parameters obtained are listed in
Table
\ref{224Ra-210Xe-parameters}, second column. 
With these parameters, the spectrum calculated is depicted in Figure \ref{Uran}.
The calculated B(E2)-transition values are listed in Table \ref{224Ra-BE2}, second (theory)
and third (experiment) column. 

As can be noted, the agreement to experiment is good and shows the effectiveness of the
pseudo-SACM to describe the collective structure of heavy nuclei.

Next, we calculated some spectroscopic factors, listed in Table \ref{Spec-U-Ra},
second column. The Equation (\ref{specfac-heavy}) was used with the approximation of the
parameter $\widetilde{B}$ as $\left(-\frac{1}{{\tilde n}_0-n_C}\right)$. The total number of relative oscillation
quanta for the system under study is ${\tilde n}_0=60$
and ${\widetilde n}_C=2$, 
thus, $\widetilde{B} \approx -0.0172$ and the exponential
factor in (\ref{specfac-heavy}) acquires the form \\ 
$e^{\widetilde{A}-0.0172({\tilde n}_0-n_c+\Delta{\tilde n}_\pi )}$
$\approx$ $(0.983)^{({\tilde n}_0-n_c+\Delta{\tilde n}_\pi )}e^{\widetilde{A}}$. 

The factor $e^{\widetilde{A}}$ is unknown and, as
explained before, the spectroscopic factors values can
be found in Table \ref{Spec-U-Ra},  
divided by $e^{{\widetilde A}}$.
As observed,
the spectroscopic factors to $\Delta {\tilde n}_\pi = 1$ are suppressed,
with a value smaller than $10^{-8}$ considered to
be zero. 

\subsection{$^{224}_{88}$Ra$_{136}$ 
$\rightarrow$ $^{210}_{82}$Pb$_{128}$
+ $^{14}_{6}$C$_{8}$}

\begin{table}[]
	\begin{center}
		\begin{tabular}{|c|c|c|c|c|c|}
		\hline
$J_i^{P_i} \rightarrow J_f^{P_f}$ &  a & 
 b & c & $U$-exp & $Ra$-exp \\ \hline
$2^+_1 \rightarrow 0^+_1$ & $251$ & $99$ & $250$ &$250 \pm 10$ & $99 \pm 3 $ \\ \hline
$2^+_2 \rightarrow 0^+_1$ & $0$ & $0.321$ & $0$ & - &  \\ \hline
$4^+_1 \rightarrow 2^+_1$ & $357$ & $141$ & $357$ & $357 
\pm 23$ &  $141 \pm 7$ \\ \hline
$4^+_2 \rightarrow 2^+_1$ & $0$ & $0.0312$ & $0$ & - & - \\ \hline
$4^-_1 \rightarrow 2^-_1$ & $309$ & $120$ & $3.554$ & - & - \\ \hline
$4^-_2 \rightarrow 2^-_1$ & $0.$  & $0.00169$ & $337$ & - & -  \\ \hline
$6^+_1 \rightarrow 4^+_1$ & $391$ & $155$  & $390$  & $385
\pm  22$ & $157 \pm 13$ \\ \hline
$2^+_1 \rightarrow 0^+_2$ & $0$ & $0.115$  & $0$  & - & -  \\ \hline
$2^+_2 \rightarrow 0^+_2$ & $269$ & $142$ & $16.13$ & - & - \\ \hline
$4^+_1 \rightarrow 2^+_2$ & $0$ & $0.00472$ & $0$ & - & - \\ \hline
$4^+_2 \rightarrow 2^+_2$ & $383$ & $58.94$ & $21$ & - & - \\ \hline
$4^-_1 \rightarrow 2^-_2$ & $0$ & $0.974$ & $0$ & - & - \\ \hline
$4^-_2 \rightarrow 2^-_2$ & $368$ & $6.25\times 10^{-6}$ 
& $0$ & - \\ \hline
		\end{tabular}
	\caption{Theoretical $B(E2)$-transition values 
	of the system a: $^{236}$U $\rightarrow$
	$^{210}$Pb + $^{26}$Ne; system b:
	$^{224}$Ra $\rightarrow$
	$^{210}$Pb + $^{14}$C and system c:
	$^{224}$U $\rightarrow$
	$^{146}$Xe + $^{90}$Sr. The last two columns list the
	experimental values for $^{236}$U and $^{224}$Ra,
	respectively.
	The unit is in WU and the
	theoretical values are
	compared to available experimental data \cite{brook}.
	Note that in $^{238}$U most of the inter-band 
	transitions are zero. This is due to the fact that the 
	ground state irrep is (54,0) and that we are working in
	the $SU(3)$ limit, thus, there are no transitions
	between states in distinct $SU(3)$ irrep.
	This changes for $^{224}$Ra, whose ground state irrep in
	the $SU(3)$ limit is (48,4), which contains several $K$
	bands ($K=0,2,4$) and allows now a transition between
	states from different bands at low energy. 
	}
	\label{224Ra-BE2}
	\end{center}
\end{table}

\begin{figure}
\centerline{
\rotatebox{270}{\resizebox{200pt}{130pt}{\includegraphics[width=0.23\textwidth]{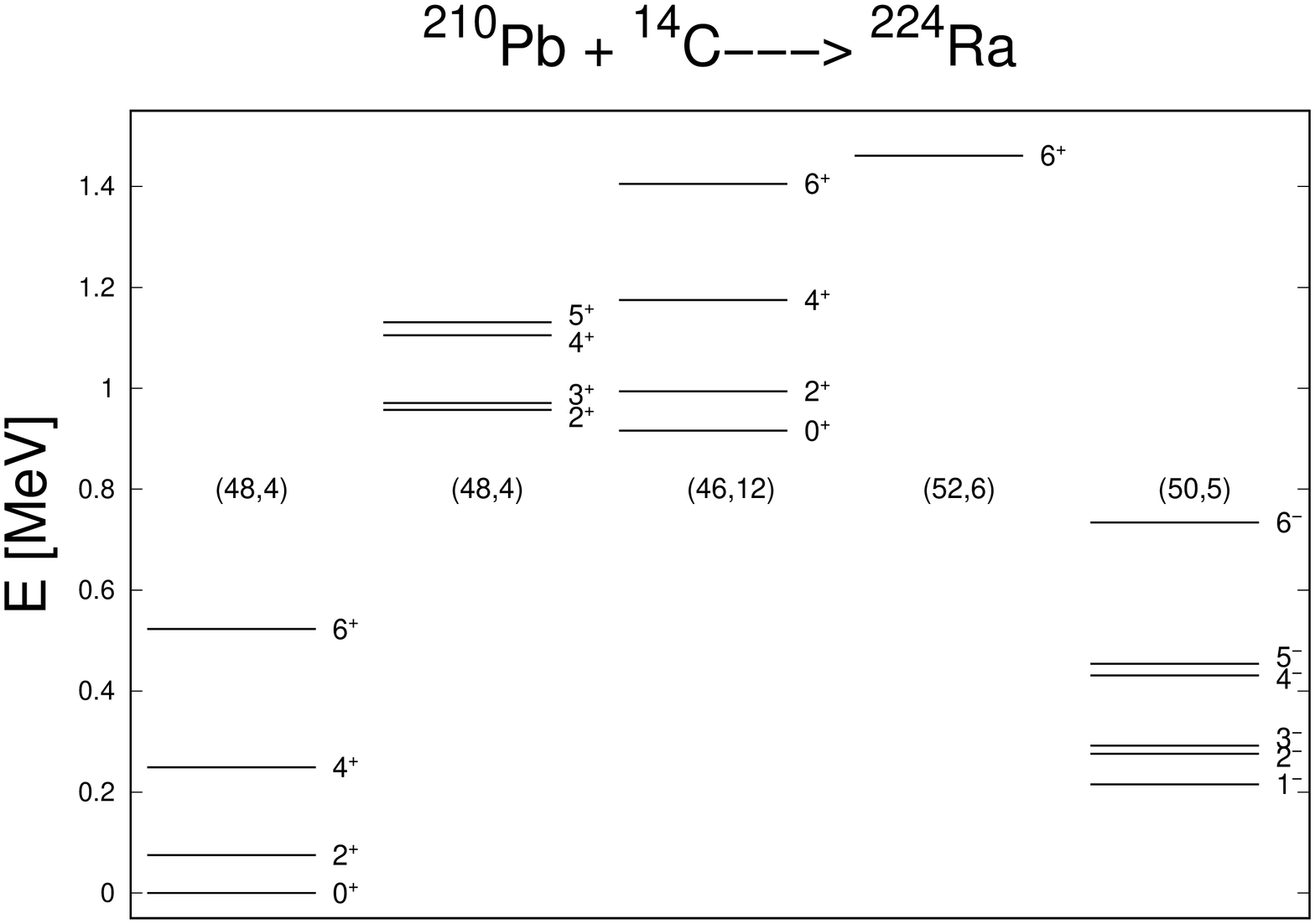}}}
\rotatebox{270}{\resizebox{200pt}{130pt}{\includegraphics[width=0.23\textwidth]{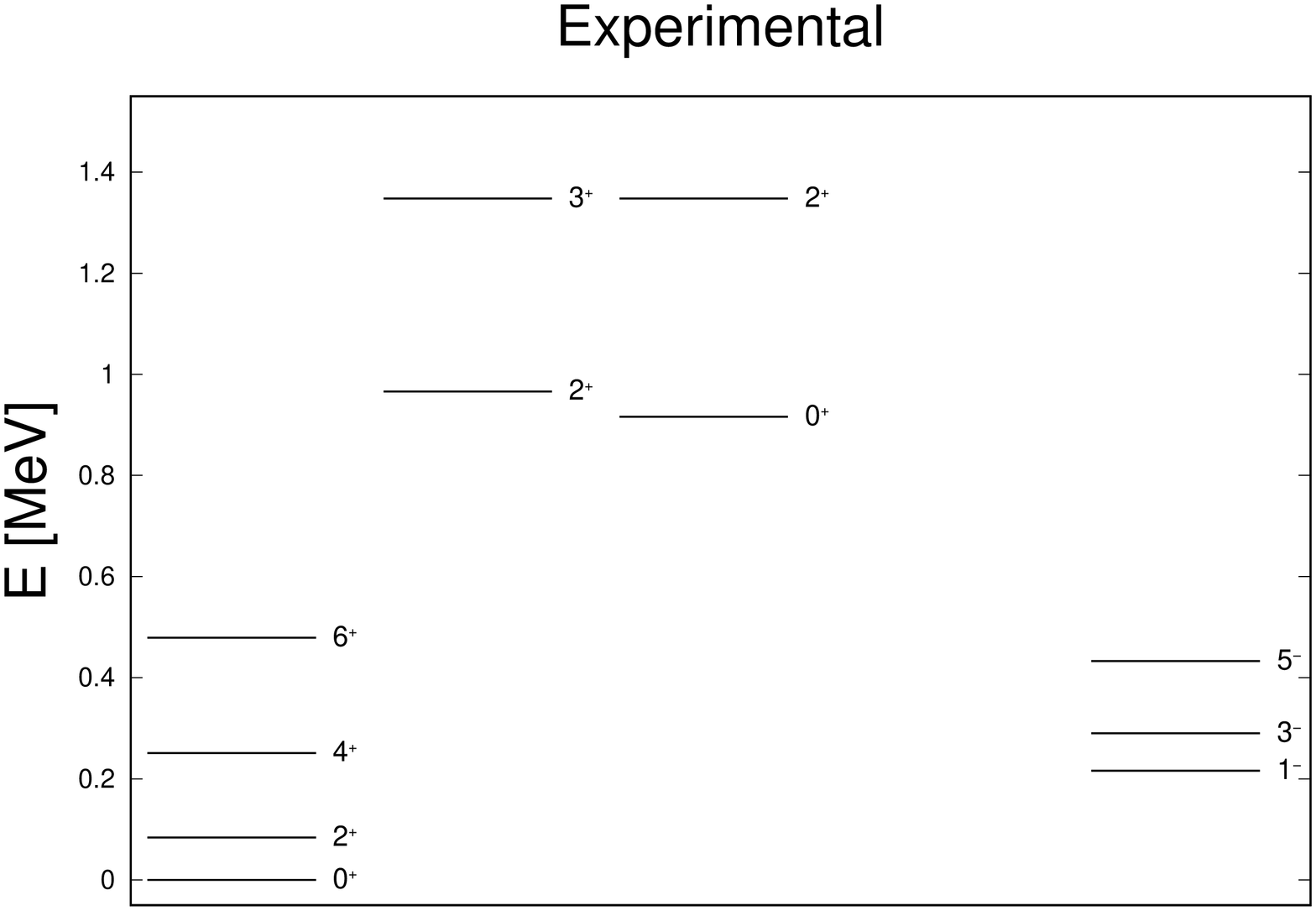}}}
}
\caption{\label{Ra} 
Spectrum of $^{224}$Ra, described by the 
clusterization $^{210}$Pb+$^{14}$C.
Only states up to angular momentum 6 are depicted 
The theoretical spectrum
(left panel) is compared to experiment (right panel).
}
\end{figure}

As in the former section, the protons and neutrons are treated separately, 
where the nucleons are filled into the Nilsson
diagram from below, at the 
deformation value $\epsilon_2 =0.150$ 
($\beta_2)=0.164$) \cite{nix-tables}.
The $\hbar\omega = 6.73$~MeV.

For $^{224}$Ra, the united nucleus, we obtain 46 protons 
in the normal orbitals and the valence shell is
${\widetilde \eta}_p=4$ with 6 valence 
protons. The $\widetilde{SU}(3)$ irrep
is $(\tilde{\lambda} , \tilde{\mu})_{p} 
=(18,0)^{\rm Ra}_p$, while for the neutrons we have   
80 normal particles with 10 in the ${\widetilde \eta}_n=5$ valence shell, 
giving $(\tilde{\lambda} , \tilde{\mu})^{\rm Ra}_{n} 
=(30,4)^{\rm Ra}_n$.
These two irreps can be coupled to the ground state irrep for $^{224}$Ra, namely 
$(\tilde{\lambda} , \tilde{\mu}) =(48,4)$.

The compilation of the valence shells, their nucleon
content and the corresponding irreps for $^{236}$U 
can be found in the former subsection.

The light cluster $^{14}$C is added on top of the heavy cluster. We have then 6 protons and 8 neutrons
in normal orbitals, which gives $(0,2)^{\rm C}_{p}$
(two holes in the $\widetilde{p}$ shell) and 
$(0,0)^{\rm C}_{n}$ (closed $\widetilde{p}$ shell).

Using the numbers just deduced, the system
$^{224}_{88}$Ra$_{136}$ $\rightarrow$ 
$^{210}_{82}$Pb$_{128}$ + $^{14}_{6}$C$_{8}$ 
can be viewed
within the $\widetilde{SU}(3)$ description
as a $^{126}_{46}\widetilde{\rm{Ru}}_{80}$ 
$ \rightarrow$ 
$^{112}_{40}\widetilde{\rm{Zr}_{72}}$
+ $^{14}_{6}\widetilde{\rm{C}}_{8}$
cluster system. 
Of course, these so-called {\it pseudo-nuclei} only
serve for illustration. In what follows, we will 
continue to use the notation for the real nuclei.

The minimal number of quanta which have to be added in the proton part of the united nucleus is 20, 
which is the result of counting the difference of the
oscillation in the united pseudo-nucleus to the sum of the
oscillation quanta of the two pseudo-clusters.  
This corresponds to a $(20,0)_{R p}$ 
irrep in the relative part. For the neutron part the
difference in the oscillation quanta is 34 and, thus, the
irrep of the relative motion is $(34,0)_{R n}$

For the proton part, the product
$\left[ (0,0)^{\rm Pb}_p \otimes (0,2)^C_p \right]$
$\otimes$ $(20,0)_{R p}$ does lead to the final 
ground state irrep $(18,0)^{\rm Ra}_p$ and,
therefore, the {\it forbiddenness} is zero.
For the neutron part, however, this is no longer the
case: The product 
$\left[ (0,0)^{\rm Pb}_n \otimes (0,0)^{\rm C}_n \right]$
$\otimes$ $(34,0)_{R n}$ does not contain the ground stte 
irrep $(30,4)^{\rm Ra}_\nu$. Applying the
formula for the {\it forbiddenness} gives $n_C^n = 2$.
These two quanta are subtracted from the relative motion,
giving $(32,0)_{R n}$ and are added to the Pb-cluster,
as one possibility (any other leads to the same final result).
The irrep used is obtained by subtracting one neutron
from the $\widetilde{\eta}_n =5$ and exciting one neutron
to the $\widetilde{\eta}_n$ = (5+2,0) = (7,0).
The neutrons in the valence shell provide the irrep
(5,0) (only one valence neutron left) and the product with
(7,0) contains the $(8,2)^{\rm Pb}_n$ irrep. The
product with the relative motion is sufficient, because
the $C$ pseudo-nucleus has the scalar neutron
irrep $(0,0)^{\rm C}_n$. The product
$(8,2)^{\rm Pb}_n \otimes (32,0)_{R n}$ contains the
ground state irrep $(30,4)^{\rm Ra}_n$.

Using the Hamiltonian in the $SU(3)$-dynamical limit, the coefficients are adjusted to 
the experimental data, listed in Table \ref{fit-U-Ra}, third column. The optimal parameters obtained are listed in
Table 
\ref{224Ra-210Xe-parameters}, third column. 
With these parameters, the spectrum calculated is depicted in Figure \ref{Ra}.
The calculated B(E2)-transition values are listed in Table \ref{224Ra-BE2}, 
third (theory) and sixth (experiment) column. 

As can be noted, the agreement to experiment is good and shows 
also in this example the effectiveness of the
pseudo-SACM to describe the collective structure of heavy nuclei.

Next, we calculated some spectroscopic factors for $^{224}$Ra, listed in Table \ref{Spec-U-Ra},
third column. The total number of relative oscillation
quanta for the system under study is ${\tilde n}_0=40$ 
($n_C=2$), 
thus $\widetilde{B} \approx -0.026$ and the exponential
factor in (\ref{specfac-heavy}) acquires the form

\beqa 
e^{\widetilde{A}-0.026({\tilde n}_0-n_c+\Delta{\tilde n}_\pi )}
& \approx & (0.974)^{({\tilde n}_0-n_c+\Delta{\tilde n}_\pi )}e^{\widetilde{A}}
~~~.
\eeqa
 
The factor $e^{\widetilde{A}}$ is unknown and thus
in Table \ref{Spec-U-Ra}, the spectroscopic factors were divided by 
$e^{{\widetilde A}}$.

\subsection{$^{236}_{92}$U$_{144}$ 
$\rightarrow$ $^{146}_{54}$Xe$_{92}$+$^{90}_{38}$Sr$_{52}$}
\label{236U-XeSr-case}

\begin{figure}
\centerline{
\rotatebox{270}{\resizebox{200pt}{130pt}{\includegraphics[width=0.23\textwidth]{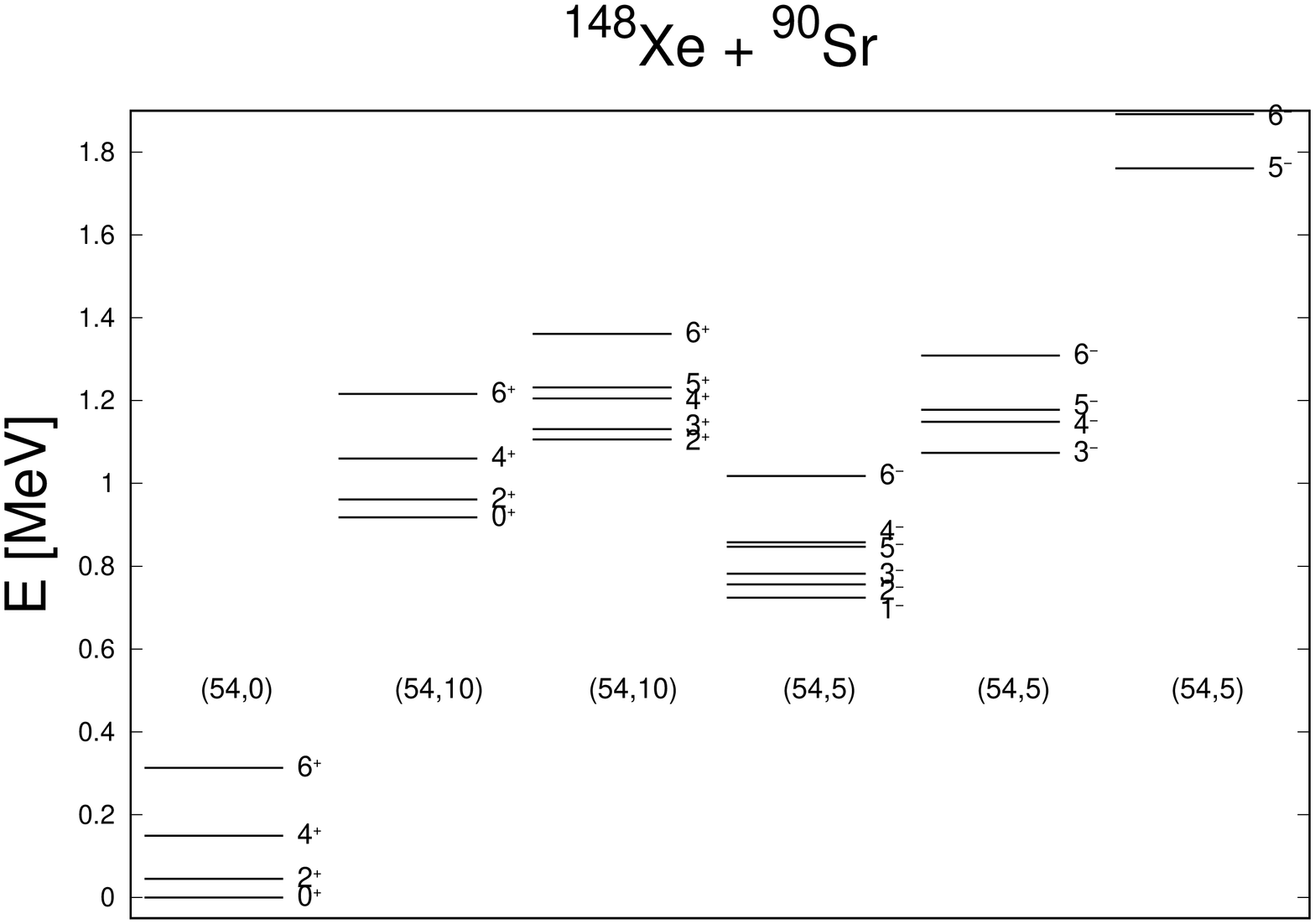}}}
\rotatebox{270}{\resizebox{200pt}{130pt}{\includegraphics[width=0.23\textwidth]{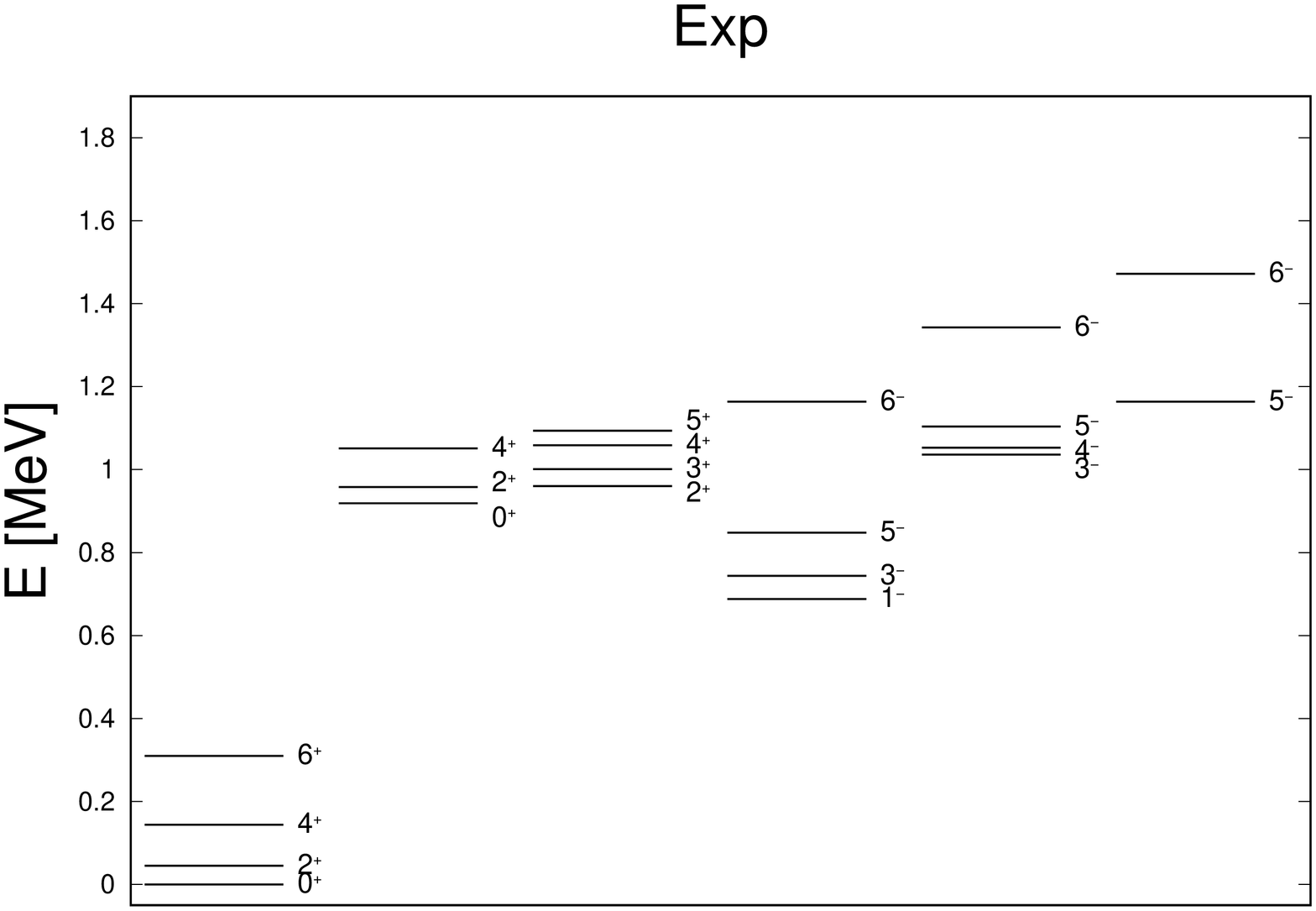}}}
}
\caption{\label{Uran-2} 
Spectrum of $^{236}$U, 
described by the clusterization $^{146}$Xe+$^{90}$Sr. 
Only states up to angular momentum are depicted.
The theoretical spectrum
(left panel) is compared to experiment (right panel).
}
\end{figure}

Consulting the table of \cite{nix-tables} the deformation of
$^{236}$U is $\epsilon_2=0.200$
($\beta_2=0.215$), further 
$\hbar\omega = 6.63$MeV. The number of normal and
unique protons and neutrons is given in subsections 
\ref{3.1}, therefore, we will resume the numbers for
$^{146}$Xe and $^{90}$Sr, only.

Counting for $^{146}$Xe 
the number of protons in the normal orbitals,
we obtain 4 in the pseudo-shell $\widetilde{\eta}_p$=3.
Taking into account the lower shells, we obtain for the
total number of protons in normal orbitals 
$\widetilde{Z}$=24. As the leading irrep,
obtained in the reduction of 
$U(\frac{1}{2}\left(\widetilde{\eta}_p +1\right)
\left(\widetilde{\eta_p} +2\right))$ $\supset$ 
$\widetilde{SU}(3)$ we have $(8,2)_p$.
For the neutron part, there are 8 neutrons in the
$\widetilde{\eta}_n$ = 4 shell, corresponding to 
$(18,4)_n$ as the leading irrep. The total
number of normal neutrons is 48.

The lighter cluster, $^{90}$Sr, is put on top of the 
$^{146}$Xe cluster. Counting the difference of the 
normal protons and neutron in the $^{236}$U nucleus 
to the $^{146}$Xe nucleus, we obtain 22 normal protons
and 34 normal neutrons for $^{90}$Sr. Distributing them 
in the $\widetilde{SU}(3)$ shell model, we have 2 valence
protons in the $\widetilde{\eta}_\pi$=3 shell and
14 neutrons in the same pseudo-shell, which correspond to 6
holes. The leading irreps are respectively
$(6,0)_p$ and $(0,12)_n$.

Counting the difference in the oscillation quanta 
in the proton sector between
the $^{236}$U nucleus and the sum of the two clusters,
we obtain 36 quanta. 
However, when we couple the two cluster
irreps and then with the relative motion irrep $(36,0)_R$,
the $(18,0)^U_p$ irrep of $^{236}$U cannot be reached,
which is a clear indication that the proton part
requires the use of the {\it forbiddenness}.
Using the formula for the {\it forbiddenness}
we obtain $n^p_C = 10$, demonstrating the importance
of the concept of {\it forbiddenness}. 
The 10 oscillation quanta
are to be added to the clusters, choosing a path
which is easier to follow:
As {\it one} possibility (all others lead to the same
result) we distribute
6 to the large and 4 to the light cluster (such that in
each an even number is added).
For the proton part, in $^{146}$Xe the irrep of one 
proton less in the valence shell is (7,1) (instead of the
former (8,2)). One proton is excited to 
$\widetilde{\eta}_p$ = (3+6,0) = (9,0). 
The product of $(7,1) \otimes (9,0)$ contains the irrep 
$(14,2)^{\rm Xe}$.
In Sr the irrep with one nucleon less in the valence shell
$\widetilde{\eta}_p = 3$ is (3,0). One nucleon is excited
the $\widetilde{\eta}_p$ = 3+4 =7, i.e., with the irrep
(7,0). In the product of $(3,0) \otimes (7,0)$ the
irrep $(10,0)^{\rm Sr}$ appears.
In the final coupling with the radial irrep 
$(36-10,0)_{p R}$ = $ (26,0)_{p R}$ we obtain
$(14,2)^{\rm Xe}_p \otimes (10,0)^{\rm Sr}_p$, which contains (4,12) and coupled with $(26,0)_{p R}$ leads
to the final proton 
irrep $(18,0)^U_p$ and, thus, we reached our
objective for the proton part

The same procedure has to be applied for the neutron part.
Counting the difference in the oscillation quanta 
in the neutron sector between
the $^{236}$U nucleus and the sum of the two clusters,
we obtain 76 quanta. However, when we couple the two cluster
irreps and then with the relative motion irrep 
$(76,0)_{n R}$,
we cannot reach the $(36,0)^U_n$ irrep of $^{236}$U,
which is a clear indication that the neutron part
requires the use of the {\it forbiddenness}.
Using the formula for the {\it forbiddenness}
we obtain $n^\nu_C = 48$, again showing that the
concept of {\it forbiddenness} is very important. 
The 48 oscillation quanta,
have to be added to the clusters.
This time, all the 48
quanta are added to the light cluster, because 
it does not matter how we distribute the $n_C^\nu$
= 48  between the clusters. This is the 
reason why we now follow this path. 
Before, in Sr there were 6 holes, now there will be
7, because one neutron is excited to the
$\widetilde{\eta}_n$ = 3+48 = 51 and the irrep which 
carries the single neutron is (51,0). The valence shell,
with 7 holes, provides the irrep (2,11).
Coupling both, $(2,11) \otimes (51,0)$, a large list is
obtained were we choose the most compact irrep, namely
(38,2). The reason to choose the most compact irrep is
that this alone leads in the product with the relative motion
to the smallest irreps possible.
Therefore, we consider the product
$\left[ (38,2) \otimes (18,4)^{\rm Xe}_n \right]$
$\otimes$ $(28,0)_{n R}$ $\rightarrow$, which leads to
the possible combination of 
$(22,14) \otimes (28,0)_{n R}$, 
where (22,14) appears in the product $(38,2) \otimes (18,4)$.
The (22,14) was chosen, because taking the product with
the relative irrep $(28,0)_{\nu R}$ it
contains the final irrep $(36,0)^{\rm U}_\nu$.

The {\it cluster irreps} for Xe and Sr are obtained
by couple linearly for each cluster the proton and 
neutron irrep. Using the above results of the compilations,
we get $(48,2)^{\rm Sr}$ and $(32,6)^{\rm Xe}$,
once coupled contains the irrep (30,30), which
coupled again with the relative motion (54,0), where 54
is the sum of the relative motions in the proton (26) and
the neutron part (28). The final result
contains the ground state
irrep of $^{236}$U, namely (54,0).

Counting only the normal nucleons, the system
$^{236}_{92}$U$_{144}$ $\rightarrow$ 
$^{146}_{46}$Xe$_{92}$ + $^{90}_{22}$Sr$_{52}$
can be viewed as the system
$^{128}_{68}$$\widetilde{{\rm Pd}}$$_{82}$ $\rightarrow$ 
$^{72}_{24}$$\widetilde{{\rm Cr}}$$_{48}$
+ $^{56}_{22}$$\widetilde{\rm Ti}$$_{34}$.
The agreement to experiment is satisfactorily, as can be
see by consulting Figure \ref{Uran-2} for the spectrum and
Table \ref{224Ra-BE2} for the transition values.
Comparing to the $^{236}$U $\rightarrow$ $^{210}$Pb
+ $^{26}$Ne, the obtained spectrum has a similar
agreement to experiment, with some shifts in the band
heads and changes in the moment of inertia.

For the calculation of the spectroscopic factor,
the parameter $\widetilde{B}$ = 
$-\frac{1}{\widetilde{n}_0-n_C}$, where
$(\widetilde{n}_0-n_C)$ = -0.0185 are the remaining total relative oscillation quanta after subtracting the {\it forbiddenness}.
The spectroscopic factors are listed in the last
column in Table \ref{Spec-U-Ra}. 

\section{Conclusions}
\label{conclusions}

We have presented an extension of the {\it Semimicroscopic Algebraic Cluster Model} (SACM),
valid for light nuclei, to the {\it pseudo-SACM}, for heavy nuclei, limiting to the
$\widetilde{SU}(3)$ dynamical symmetry limit. 
Though, there exist earlier attempts to extend the
SACM to heavy nuclei, we found it necessary to construct a model, which enables us to 
determine the complete spectrum and
circumvent some conceptual and practical problems of the former approaches and to deliver a consistent
procedure, as working in the same mean field of the parent nucleus and
simultaneously conserving the simplicity of the SACM
for light nuclei.

Protons and neutrons have to be treated separately, because they occupy different shells.
Only at the end they are coupled together.
The protons and neutrons are 
distributed within the normal and
unique orbitals in such a manner that the sum of normal nucleons of the clusters
is the same as in the parent nucleus. The construction of the model space is in complete analogy 
to the SACM.

As examples, we considered 
$^{236}$U $ \rightarrow$ $^{210}$Pb+$^{26}$Ne,
$^{224}$Ra $\rightarrow$ $^{210}$Pb+$^{14}$C
and
$^{236}$U $\rightarrow$ $^{146}$Xe+$^{90}$Sr. 
We demonstrated that the model is
able to describe the spectrum and electromagnetic transition probabilities. 
Spectroscopic factors were also calculated, without further fitting and they can be
considered as a prediction of the model. 
With this, we demonstrated the usefulness of the pseudo-SACM
for treating heavy nuclei. A more systematic study of
several nuclei is planned in the future.

The restriction to the $\widetilde{SU}(3)$ dynamical symmetry limit has to be relaxed
in future applications, including the other 
dynamical symmetries as $SO(4)$. One has to study the extension of $\widetilde{SU}(3)$ too,
including the active participation of nucleons in the unique orbitals.
Also the study of phase transitions is of
interest, requiring the use of the geometrical mapping 
\cite{geom} of the SACM.

\section*{Acknowledgments}
We acknowledge financial support form DGAPA-PAPIIT 
(IN100421).


\vskip 1cm
\begin{thebibliography}{00}

\bibitem{cseh-letter} J. Cseh, Phys. Lett. B \textbf{281} (1992), 173.

\bibitem{cseh-levai-anph}
 J. Cseh and G. L\'evai, Ann. Phys. (N.Y.) \textbf{230}, 165
(1994).

\bibitem{scheid1995} W. Greiner, J. Y. Park and W Scheid,
{\it Nuclear Moelcules}, (World Scientific, Singapure, 1995).

\bibitem{hess-1984} P. O. Hess and W. Greiner, Il Nuovo
Cimento {\bf 83}, 76 (1984).

\bibitem{sacm-appl1}H.  Y\'epez-Mart\ii nez, 
M. J. Ermamatov,
P. R. Fraser and P. O. Hess, Phys. Rev. C {'bf 86} (2012),
034309.

\bibitem{cseh-algora} A. Algora and J. Cseh, J. Phys. G
{\bf 22}, L39 (1996)

\bibitem{hecht} K.T. Hecht and A. Adler, Nucl. Phys. 
A {\bf 137}, 129 (1969)

\bibitem{arima} A. Arima, M. Harvey and K. Shimizu, Phys. Lett. B {\bf 30}, 517 (1969)

\bibitem{cseh-scheid} J. Cseh, R. K. Gupta and W. Scheid,
Phys. Lett. B {\bf 299}, 205 (1993)

\bibitem{hunyadi} P. O. Hess, A. Algora, M. Hunyadi, J. Cseh,  Eur. Phys. Jour. {\bf A15}, 449 (2002)

\bibitem{ring} P. Ring and P. Schuck, {\it The Nuclear
Many-Body Problem}, (Springer, Heidelberg,1980).

\bibitem{sacm-fission1} A. Algora, J. Cseh and P. O. Hess,
J. Phys. G {\bf 24}, 2111 (1998)

\bibitem{sacm-fission2} A. Algora, J. Cseh and P. O. Hess,
J. Phys. G {\bf 25}, 775 (1999)

\bibitem{sacm-fission3} A. Algora, J. Cseh, J. Darai and 
P. O. Hess, Phys. Lett. B {\bf 639}, 451 (2006)

\bibitem{hess-86} H. Y\'epez-Mart\'inez, M. J. Ermamatov, P. R. Fraser and P. O. Hess, 
Phys. Rev. C {\bf 86}, 034309 (2012)

\bibitem{phase-I} H. Y\'epez-Mart\'inez, P. R. Fraser, P. O. Hess and
G. L\'evai, Phys. Rev. C {\bf 85}, 014316 (2012)

\bibitem{phase-II} P. R. Fraser, H. Y\'epez-Mart\'inez, P. O. Hess and
G. L\'evai, Phys. Rev. C {\bf 85}, 014317 (2012)

\bibitem{david} D. Lohr-Robles, E. L\'opez-Moreno 
and P. O. Hess, Nucl. Phys. A {\bf 992}
(2019), 121629. 

\bibitem{gilmore} R. Gilmore, {\it Catastrophe
Theory for Scientists and Enegineers}, (Wiley, New York, 1981)

\bibitem{renorm} H. Y\'epez-Mart\ii nez, G. E. Morales-Hern\'andez, P. O. Hess, G. L\'evai and P. R. Fraser,
Int. J. Mod. Phys. E {\bf 22}, 1350022 (2013)

\bibitem{cseh2020} J. Cseh, Phys. Rev. C {\bf 101} (2020), 
054306.

\bibitem{arima-quart} A. Arima, V. Gillet, and J. Ginocchio, 
Phys. Rev. Lett. {\bf 25} (1970), 1043.

\bibitem{danos-quart} M. Danos and V. Gillet, 
Phys. Rev. {\bf 161} (1967), 1034.

\bibitem{cseh-quart} J. Cseh, Phys. Lett. B {\bf 743} 
(2015), 213.

\bibitem{bonatsos2017} D. Bonatsos,  I. E. Assimakis, 
N. Minkov, et al., Phys. Rev. C 
{\bf 95} (2017), 064325.

\bibitem{multi} J. Cseh, Phys. Rev. C {\bf 50} (1994), 2240.

\bibitem{cseh2006} A. Algora, J. Cseh, J. Darai and P.O. Hess, Phys. Lett. B {\bf 639}, 451 (2006)

\bibitem{ring} P. Ring and P. Schuck, {\it The Nuclear
Many-Body Problem}, (Springer, Heidelberg, 1980).

\bibitem{plb1994} O. Casta\~nos, V. Vel\'azquez, P. O. Hess and J. G. Hirsch, Phys. Lett. B {\bf 321} (1994), 303.

\bibitem{fieldsu3} W. Greiner and J. A. Maruhn,
{\it Nuclear Models}, (Springer, Berlin-Heidelberg, 1996).

\bibitem{draayer-book} J. Draayer, {\it Fermion models}, 
in {\it Algebraic Approaches to
Nuclear Structure}, ed. R. Casten et al. 
(Harwood Academic Publisher, Pennsylvania, 1993) p. 423.

\bibitem{NPA1994} D. Troltenier, J. P. Draayer, P. O. Hess and O. Casta\~nos, Nucl. Phys. A {\bf 576} (1994), 351.

\bibitem{smirnov} Yu. F. Smirnov and Yu. M. Tchuvil'sky, Phys. Lett. B {\bf 134}, 25 (1984)

\bibitem{wildermuth}  K. Wildermuth and Y. C. Tang, \emph{A Unified Theory of
the Nucleus} (Friedr. Vieweg \& Sohn Verlagsgesselschaft mbH, Braunschweig,
1977).

\bibitem{huitz-2015} H. Y\'epez-Mart\ii nez, P. O. Hess, 
J. Phys. G {\bf 42}, 095109 (2015)

\bibitem{pseudo-sympl} O.Casta\~nos, P.O.Hess, P.Rocheford, J.P.Draayer, Nucl. Phys. {\bf A524}, 469 (1991)

\bibitem{bahri} C. Bahri, D. J. Rowe and J. P. Draayer,
Comput. Phys. Commun. {\bf 159} (2004), 121.

\bibitem{nix-tables} P. M\"oller, J.R. Nix, W.D. Myers, W.J. Swiatecki, At. Data Nucl. Data Tables 
{\bf 59}, 185 (1995)

\bibitem{draayer1} O. Casta\~nos, J.P. Draayer and Y. Leschber, Ann. of Phys. {\bf 180}, 290 (1987)

\bibitem{NPA576} D. Troltenier, J. P. Draayer, O. Casta\~nos
and P. O. Hess, Nucl. Phys. A {\bf 576} (1994), 351.

\bibitem{eisenberg} J. M. Eisenberg and W. Greiner, {\it Nuclear Theory I: 
Nuclear Models}, 3rd edn (Amsterdam, North-Holland, 1987)

\bibitem{orient} J. Cseh, J. Darai, A. Algora, H. Y\'epez-Mart\ii nez, P. O. Hess,
Rev. Mex. F\ii s. {\bf 54} (S3), 30 (2008)

\bibitem{rowe1} D. J. Rowe, Rep. Progr. Phys. {\bf 48}, 1419 (1985)

\bibitem{TE} Castaños O., Draayer J. P.,
Leschber Y.; Zeitschr. f. Physik A  
{\bf 329} (1988), 33.

\bibitem{hw} J. Blomqvist and A. Molinari, Nucl. Phys. A \textbf{106}, 545 (1968)

\bibitem{escher} J. Escher and J.P. Draayer, 
J. Math. Phys. 39, 5123
(1998)

\bibitem{fraser-2012} H. Y\'epez-Mart\ii nez, 
M. J. Ermamatov, P. R. Fraser and P. O. Hess,
Phys. Rev C {\bf 86} (2012), 034309.

\bibitem{specfac-draayer} P. O. Hess, A. Algora, J. Cseh and J. P. Draayer,  Phys. Rev. {\bf C70}, 
051303(R) (2004)

\bibitem{draayer2} J. P. Draayer, Nucl. Phys. 
A {\bf 237}, 157 (1975)

\bibitem{geom} P. O. Hess, G. L\'evai and J. Cseh, Phys. Rec C {\bf 54}, 2345 (1996)

\bibitem{brook} National Nuclear Data Center, 
http://www.nndc.bnl.gov 
 2111.

\end{thebibliography}
\end{document}